\documentclass[apj]{emulateapj}

\usepackage{graphicx}
\usepackage{lscape}

\slugcomment{accepted to the Astrophysical Journal}

\begin{document}

\title{The Evolution of Galaxy Mergers and Morphology \\ at $z < 1.2$ in the Extended Groth Strip}
\author{Jennifer M. Lotz\altaffilmark{1,2}, M. Davis\altaffilmark{3}, S.M. Faber\altaffilmark{4}, 
P. Guhathakurta\altaffilmark{4}, S. Gwyn\altaffilmark{5}, J. Huang\altaffilmark{6},  D.C. Koo\altaffilmark{4},\\
E. Le Floc'h\altaffilmark{7,8,9}, Lihwai Lin\altaffilmark{4}, J. Newman\altaffilmark{10, 11} K. Noeske\altaffilmark{4}, 
C. Papovich\altaffilmark{9,12}, C.N.A. Willmer\altaffilmark{12}, A. Coil\altaffilmark{11, 12}, \\
C. J. Conselice\altaffilmark{13}, M. Cooper\altaffilmark{3}, A. M. Hopkins\altaffilmark{14}, A. Metevier\altaffilmark{4, 15},
J. Primack\altaffilmark{16}, G. Rieke\altaffilmark{12}, B. J. Weiner\altaffilmark{12}}
\altaffiltext{1}{National Optical Astronomical Observatories, 950 N. Cherry Avenue, Tucson, AZ 85719, USA; lotz@noao.edu}
\altaffiltext{2}{Leo Goldberg Fellow}
\altaffiltext{3}{Department of Astronomy, University of California, Berkeley, CA, USA}
\altaffiltext{4}{UCO/Lick Observatory, Department of Astronomy \& Astrophysics, University of California, Santa Cruz, USA}
\altaffiltext{5}{Department of Physics and Astronomy, University of Victoria, Victoria, BC, Canada}
\altaffiltext{6}{Harvard-Smithsonian Center for Astrophysics, Cambridge, MA, USA}
\altaffiltext{7}{Institute for Astronomy, University of Hawaii, Honolulu, HI, USA}
\altaffiltext{8}{Observatoire de Paris, Meudon, France}
\altaffiltext{9}{Spitzer Fellow}
\altaffiltext{10}{Lawrence Berkeley National Laboratory, Livermore, CA, USA}
\altaffiltext{11}{Hubble Fellow}
\altaffiltext{12}{Steward Observatory, University of Arizona, Tucson, AZ, USA}
\altaffiltext{13}{School of Physics and Astronomy, University of Nottingham, Nottingham, U.K.}
\altaffiltext{14}{School of Physics, University of Sydney, Sydney, Australia}
\altaffiltext{15}{NSF Astronomy and Astrophysics Postdoctoral Fellow}
\altaffiltext{16}{Department of Physics, University of California, Santa Cruz, USA}

\begin{abstract}
We present the quantitative rest-frame $B$ morphological evolution and galaxy merger fractions at 
$0.2 < z < 1.2$ as observed by the All-wavelength Extended Groth Strip
International Survey (AEGIS). We use the Gini coefficent and $M_{20}$ to identify major mergers 
and classify galaxy morphology for a volume-limited sample of 3009 galaxies brighter than $0.4 L_B^{*}$, 
assuming pure luminosity evolution of 1.3 $M_{B}$ per unit redshift.  
We find that the merger fraction  remains roughly constant at $10 \pm 2$\% 
for $0.2 < z < 1.2$.  The fraction of E/S0/Sa increases from $21\pm 3$\% 
at $z \sim 1.1$ to  $44 \pm 9$\% at $z \sim 0.3$, while the fraction of Sb-Ir decreases from  $64 \pm 6$\% 
at $z \sim 1.1$ to $47 \pm 9$\% at $z \sim 0.3$.  The majority of $z < 1.2 $ {\it Spitzer} 
MIPS 24 $\mu$m sources with L(IR) $> 10^{11} L_{\odot}$ are disk galaxies, and only $\sim$ 15\% are classified
as major merger candidates.  Edge-on and dusty disk galaxies
(Sb-Ir) are almost a third of the red sequence at $z \sim 1.1$, while E/S0/Sa make
up over 90\% of the red sequence at $z \sim 0.3$.  Approximately 2\% 
of our full sample are red mergers. We conclude (1) the merger rate does not evolve strongly between $0.2 < z < 1.2$; 
(2) the decrease in the volume-averaged
star-formation rate density since $z \sim 1$ is a result of declining star-formation in disk galaxies rather than
a disappearing population of major mergers; (3) the build-up of the red sequence at $z < 1$
can be explained by a doubling in the number of spheroidal galaxies since $z \sim 1.2$. 
\end{abstract}


\section{INTRODUCTION}

Although the evidence for a cold dark-matter dominated universe in which structure
grows hierarchically is overwhelming (e.g. Spergel et al. 2003, 2007), 
the role of mergers in galaxy assembly and star-formation remains unclear.
Simulations of major mergers (Mihos \& Hernquist 1996; Cox et al. 2006), and observations of
local gas-rich mergers (e.g., the Antennae; Schweizer 1982) indicate 
that the merger process can trigger violent starbursts and transform
disks into spheroids.  The correlation between galaxy morphology and color (e.g., 
de Vaucouleurs 1961; Blanton et al. 2003)
suggests that a galaxy's star-formation history is closely tied to its morphological
evolution.  However, recently it has been recognized that there may be different timescales for 
the formation of stars and the assembly of those stars into massive spheroidals (e.g., 
Kauffmann \& Charlot 1998; De Lucia et al. 2006). Moreover, mergers may be not the
only way to produce spheroids -- galaxy harassment, virial shock heating, 
and explosive feedback from active galactic nuclei could all result in spheroidal 
galaxies (e.g. Moore, Lake, \& Katz et al. 1998, Birnboim, Dekel \& Neistein 2007; Hopkins et al. 2006).   
Tracking the galaxy merger rate and the
evolution of galaxy morphology as a function of redshift and color can constrain
the contribution of mergers to the formation of stars and spheroidal systems.
  
The galaxy merger rate is estimated from the number density of morphologically 
disturbed galaxies or kinematically close pairs.  Accurate merger rates from
kinematic pair statistics are challenging, requiring spectroscopic velocities for both companions, 
and are therefore observationally expensive and suffer from incompleteness.  
Recent work has attempted to derive the close pair statistics from the two-point correlation
function of large numbers of galaxies (Masjedi et al. 2006; Bell et al. 2006b), which
also requires high quality redshifts and suffers from sampling problems at very small
scales.
Morphology studies, especially with deep and high resolution Hubble Space Telescope ($ HST$) images, 
do not suffer from the same incompleteness problems, yet
the morphological merger fraction is also elusive.
Morphological disturbances which are the result of recent or on-going mergers can
be determined via visual classifications or by identifying outliers in quantitative
morphology distributions.  The visual classification of
galaxy morphology is time-consuming and subjective.  Also, not all visually `peculiar' galaxies are necessarily
major mergers (e.g. some are irregulars) and high-redshift disks are more likely to be mis-classified
as peculiars due to surface-brightness effects and wavelength-dependent morphologies 
(Brinchmann et al. 1998; Marcum et al. 2001; Windhorst et al. 2002; 
Papovich et al. 2003; Kampczyk et al. 2007).  Automated morphological analysis is a promising technique for 
identifying mergers that has developed in recent years (e.g. Abraham et al. 1996, Conselice et al. 2003, 
Lotz et al. 2004, Scarlata et al. 2007a). Most previous quantitative
morphology studies aimed at finding merger candidates have used concentration and asymmetry measurements 
(e.g. Abraham et al. 1996; Conselice et al. 2003; Cassata et al. 2005; Menanteau et al. 2006). 
However, asymmetry measures are less sensitive to merger remnants in late stages, and require high 
signal-to-noise images to overcome noisy background residuals.   In this paper, we use the 
Gini coefficient ($G$) and the second-order 
moment of the brightest 20\% of a galaxy's pixels ($M_{20}$) which has been shown to be more effective 
than concentration and asymmetry at identifying late-stage mergers and classifying galaxies
(Lotz, Primack, \& Madau 2004, hereafter LPM04). 
The Gini coefficient and $M_{20}$ are also more robust at low signal-to-noise ratios 
(LPM04; Lotz et al. 2006).   Recent attempts
to combine $G$, $M_{20}$, $C$, and $A$ using a principal component analysis also have yielded
promising results (Scarlata et al. 2007a). 

The initial results of $z \sim 1$ morphological studies revealed a population
of visually disturbed, star-forming galaxies (Driver et al. 1995; Glazebrook et al. 1995; Abraham et. al 1996).  
This discovery led to the claim that the increase in the volume-averaged star-formation rate density
from $z=0$ to $z>1$ (e.g., Lilly et al. 1996; Madau et al. 1996; Hopkins 2004) was
partially the result of a dramatic increase in the merger rate at higher redshift 
(Le F\`{e}vre et al. 2000; Bridge et al. 2007).  Pair count studies also indicated 
strong evolution in the number of close pairs at $0 < z < 1$ (Patton et. al. 2002; Le F\`{e}vre et al. 2000
but see Carlberg et al. 2000 and Lin et al. 2004).
However, these results suffered from cosmic variance and small-number statistics (typically tens of
merger candidates in a sample of a few hundred). 

With recent {\it Spitzer Space Telescope} infrared surveys and 
high-resolution Hubble Space Telescope ($HST$) morphologies, the conclusion that high-redshift
star formation is driven by mergers has been re-examined.
Infrared luminous galaxies dominate the star-formation density at $z \sim 1$ (Le Floc'h
et al. 2005), but the trigger for their star-formation remains strongly debated.  
The majority of these galaxies at $z \sim 0.7$ have been visually classified as spirals, not
peculiars or mergers (Flores et al. 1999; Bell et al. 2005; Melbourne, Koo, \& Le Floc'h 2005).  However, 
these galaxies have higher asymmetries than normal spirals
(Shi et al. 2006; Bridge et al. 2007) and may be more likely to be found in interacting close pairs
(Lin et al. 2007; Bridge et al. 2007). 

Another question the merger rate may address is the formation of red spheroidal galaxies.
Recent studies have shown that the color distribution of galaxies out to $z \sim 1.2$ is
bi-modal, and that the number density of luminous red galaxies has increased by a factor 
of 2-4 since $z \sim 1$ (Bell et al. 2004a; Faber et al. 2007; Brown et al. 2007), implying a large influx
of red galaxies.  The majority of red galaxies at
$z \leq 0.7$ are spheroid-dominated systems (Blanton et al. 2003; Bell et al. 2004b; Weiner et al. 2005;
Scarlata et al. 2007b). 
Several mechanisms may be responsible for moving galaxies
onto and along the red sequence, including merging of massive gas-rich galaxies, 
the ``quenching'' of disk galaxies via gas-stripping or AGN feedback, and the merging
of red galaxies (Bell et al. 2004a; Faber et al. 2007).  
A number of low redshift E/S0 show residual disturbances (e.g., van Dokkum 2005; Schweizer et al. 1990), 
suggesting that many have experienced mergers in previous Gyr.
However, the amount of stellar mass in massive spheroids appears to have changed very little
since $z \sim 1$ (Bundy et al. 2005; Brinchmann \& Ellis 2000), although there is
evidence that less massive early-types may have assembled more recently (Bundy et al. 2005; 
Treu et al. 2005; Papovich et al. 2006).

In this paper, we apply a new method for identifying major merger candidates to the
recent $HST$ Survey of the Extended Groth Strip (EGS), which is part of the 
All-Wavelength Extended Groth strip International Survey (AEGIS) 
collaboration (Davis et al. 2007).  We use the Gini coefficient $G$,
which is a measure of the relative distribution of galaxy pixel flux values, and
$M_{20}$, the relative second-order moment of the brightest 20\% of a galaxy's
pixels, to classify galaxies as major mergers, E/S0/Sa, or Sb/Sc/Ir
out to $z \sim 1.2$.  In our previous work, we have demonstrated that $G$ and $M_{20}$ are
robust morphological classifiers for galaxies as distant as $z \sim 4$ (Lotz et al. 2006),
and are an effective way to find major merger candidates (LPM04). 
Using spectroscopic redshifts from the DEEP2 survey
and photometric redshifts from the Canada-France-Hawaii Telescope Legacy Survey (CFHTLS), 
we construct a volume-limited sample of 3009 galaxies 
brighter than $0.4 L_B^*$ at $0.2 < z < 1.2.$, assuming pure luminosity evolution
of 1.3 $M_B$ per unit redshift (Faber et al. 2007). 
We identify 312 merger candidates brighter than $0.4 L^{*}_{B}$ and find that 
the fraction of major merger candidates is roughly constant at $\sim$ 10\% for $0.2 < z < 1.2$ for both
samples.  We discuss the evolution of morphology for the bright galaxy population, luminous infrared galaxies, 
and the red sequence, and the implications for the build up of red spheroidals at $z < 1$. 
Throughout this paper, we use the term ``morphologically disturbed'' to refer to objects whose
visual or quantitative morphologies classify them as merger candidates.  We use the
term ``morphological merger fraction'' and ``fraction of morphologically disturbed objects''
interchangeably to refer to the fraction of merger candidates identified via their morphologies
as opposed to the merger fraction determined by close pair statistics. 
We adopt a
$H_0 = 70$, $\Omega_\lambda = 0.7$, $\Omega_m = 0.3$ cosmology throughout this work. 

\section{Extended Groth Strip Observations}
\begin{figure*}
\plotone{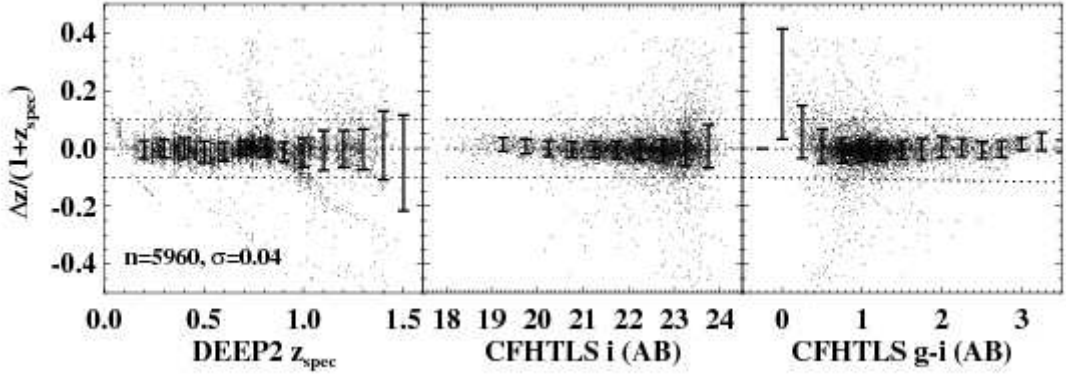}
\caption{ DEEP2 EGS spectroscopic redshifts, CFHTLS $i$ magnitude, and CFHTLS $g-i$
color v. CFHTLS photometric redshift errors [$(z_{phot} - z_{spec})/(1 + z_{spec})$] 
for 5960 objects matched to quality $\ge 3$ DEEP2 spectroscopic redshifts.  
The error-bars show the standard deviation in $\Delta z/ (1+z_{spec})$ 
as a function of redshift, magnitude, and color.  The dotted lines show $\pm$ 3 $\sigma$ for the full 
sample ($= 0.117$). }
\end{figure*}

\subsection{$HST$ ACS imaging and morphologies}
Deep $HST$ images of the EGS were
taken with the Advanced Camera for Surveys (ACS) as part of GO
Program 10134 (PI M. Davis; Davis et al. 2007) and the AEGIS collaboration.   
The EGS was imaged in $V$ (F606W, 2260s)
and $I$ (F814W, 2100s) over 63 tiles in a $\sim$ 10.1\arcmin\ $\times$ 70.5\arcmin\ strip 
centered at J2000 RA= 14h19m18s, Dec=+52$\degr$ 49$\arcmin$ 25$\arcsec$.   
Each tile was observed with 4 pointings per filter and combined with the 
STSDAS.multidrizzle package using a square kernel.  The final images have a 
pixel scale of $0.03 \arcsec$ per pixel and PSF of $0.12 \arcsec$ FWHM.
The drizzling procedure results in correlated pixels in the output image; therefore
the true sky noise for object photometry and morphology 
was calculated from the weight maps output by multidrizzle.   
The 5-$\sigma$ limiting magnitudes for a point source 
are $V_{F606W} = 28.14$ (AB) and $I_{F814W} = 27.52$ (AB)  
within a circular aperture of radius $0.12 \arcsec$ ($\sim$ 50 pixels). 
For an extended object, the 5-$\sigma$ limiting magnitudes are $V_{F606W} = 26.23$ (AB) 
and $I_{F814W} = 25.61$ (AB) assuming a circular aperture of radius $0.3\arcsec$ ($\sim 
314$ pixels). Hereafter all references to ACS photometry in $V_{F606W}$ and $I_{F814W}$ 
are in the AB system.

We detected objects in summed ACS $V+I$ images and constructed initial galaxy 
segmentation maps using the SExtractor galaxy photometry software
(Bertin \& Arnouts 1996) and a detection threshold of $1.5 \sigma$ and 50 pixels. 
These detection maps and the ACS zeropoints (Sirianni et al. 2005) were applied to each 
band separately to create the ACS photometric catalogs.  ACS isophotal magnitudes
are known to suffer from significant aperture corrections (Sirianni et al. 2005), 
therefore we use SExtractor `BEST' magnitudes which attempt to correct for missing 
light and are in better agreement with ground-based magnitudes from DEEP and CFHTLS.  

We selected all non-stellar objects with 
SExtractor CLASS\_STAR $< 0.9$ and $I_{F814W} < 25.0$ that did not lie 
within 50 pixels of a tile edge for our automated morphology analysis, covering an effective 
area of 710.9 arcmin$^{2}$ in the ACS images. For each galaxy
the automated morphology analysis code uses the SExtractor segmentation
map to mask out background and foreground objects. Then it calculates the Petrosian radius $r_p$ 
and assigns pixels brighter than the surface brightness at $r_p$ to the galaxy, 
and computes $G$ and $M_{20}$ using these pixels (see Lotz et al. 2006 
for details of the morphology analysis code). 
Out of 18831 galaxies brighter than 
$I_{F814W} = 25.0$  more than 50 pixels from the image edge, 
we obtained morphologies for 10832 ($V$) and 11465 ($I$) EGS galaxies with 
$\langle S/N \rangle$ per pixel $\ge 2.5$, $r_p \ge 0.3 
\arcsec\ $, and contiguous segmentation maps  (Figure 1). About 3\% of all measured
$I < 25$ objects were 
flagged as having non-continuous segmentation maps, usually the result of bad 
foreground/background object masks.  Those flagged objects which meet our final
sample criteria (\S 2.4) were reanalyzed by hand to properly mask out foreground/background
objects. Another 6\% had Petrosian radii $r_p <0.3 \arcsec$
and were too compact to derive morphologies. The majority of low $\langle S/N \rangle$ per pixel 
objects are low-surface brightness galaxies with $\mu_{V} > 24.7$ or $\mu_I > 24.2$
(see Section \S 2.4).

\begin{figure*}
\epsscale{2.0}
\plottwo{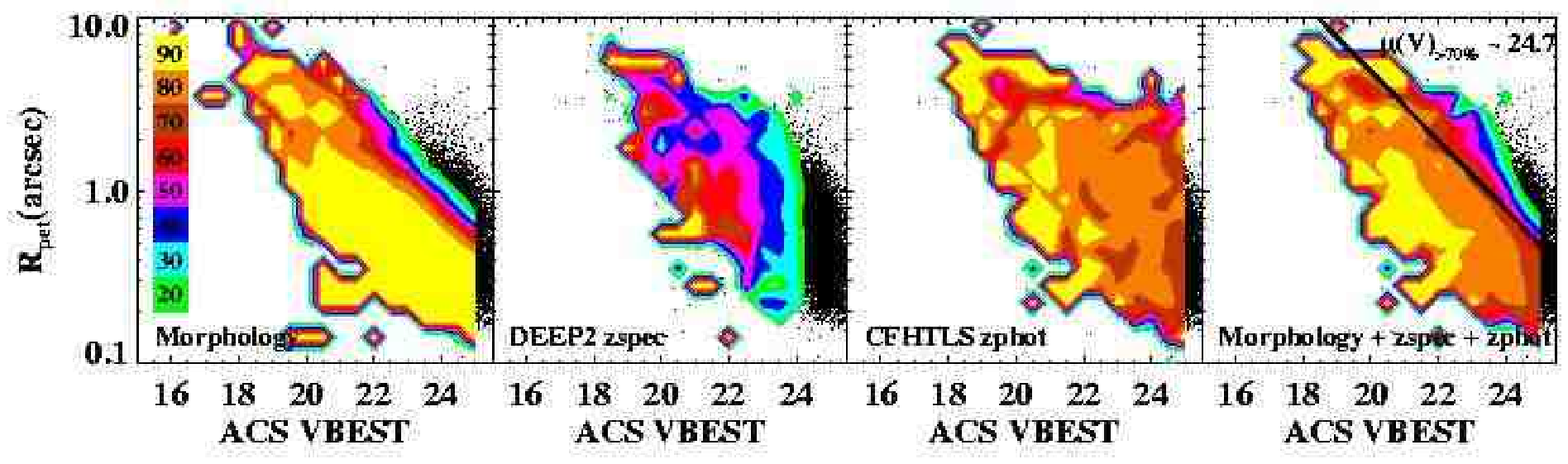}{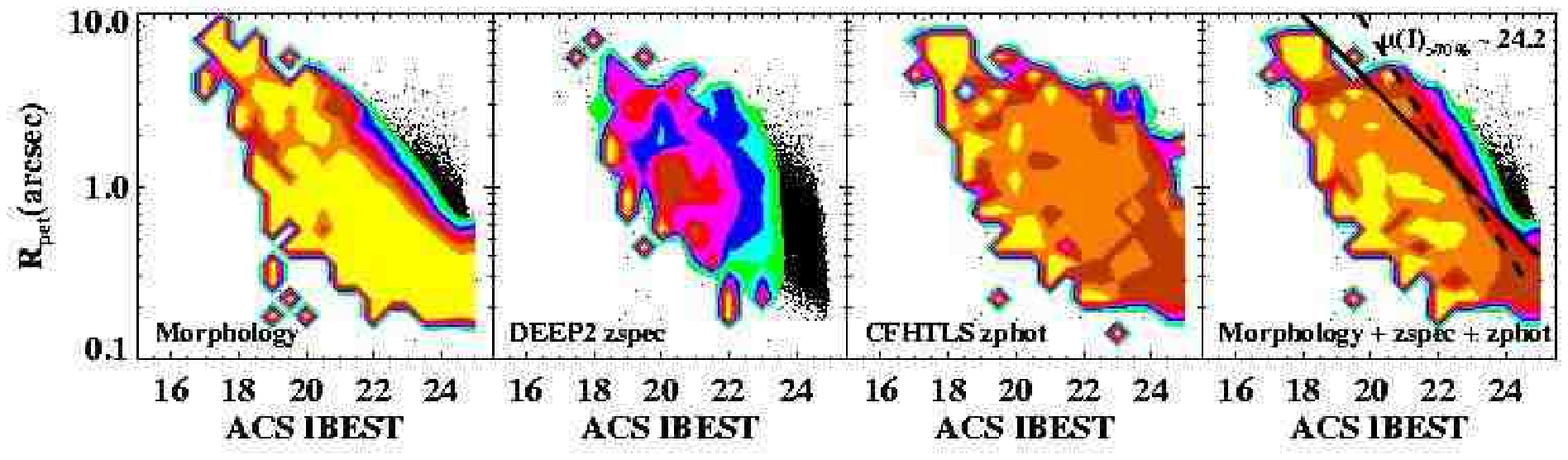}
\caption{Completeness v. Petrosian radius and apparent magnitude for the morphology and redshift catalogs. The
black dots are all ACS galaxy detections brighter than $I =25$.  The contours give the completeness in
10\% intervals, from 20\% (green) to 90\% (yellow).The solid lines in the right-hand panels
are for $\mu$(V) $\sim$ 24.7 magnitudes per sq. arcsec (top right-hand panel) and $\mu$(I) $\sim$ 24.2 
magnitudes per sq. arcsec (bottom right-hand panel) 
where the average signal-to-noise = 2.5 within R$_{pet}$ for a face-on galaxy.  The dashed line in 
bottom right-hand panel shows the 70\% completeness limit (brown contours) for the joint morphology-redshift catalog.}
\end{figure*}

\begin{figure*}
\epsscale{1.0}
\plotone{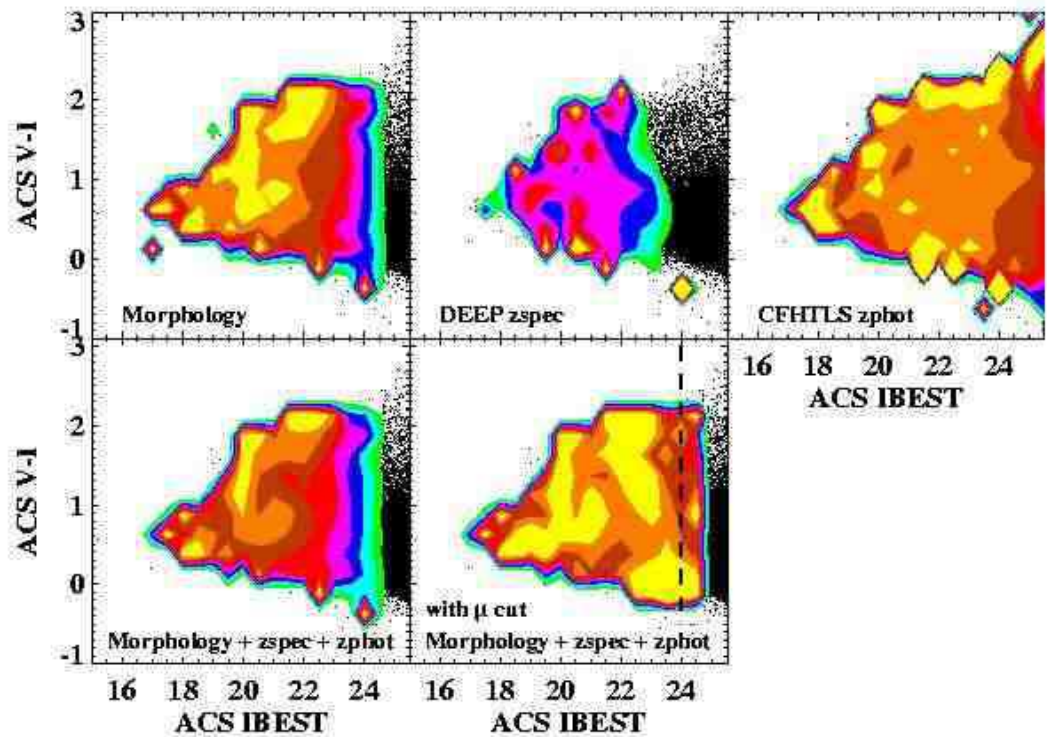}
\caption{Completeness v. observed ACS $V-I$ color and $I$ magnitude for the morphology and redshift catalogs.  
The completeness contour levels  are the same as Figure 2. The last panel
shows the completeness for objects within the 70\% completeness cut given by the dashed line in
the last panel of Figure 2.}
\end{figure*}
\subsection{Redshifts}
The EGS is one of four DEEP2 Galaxy Redshift Survey fields -- see Davis et al. 2002, Coil et al. 2004, 
Davis et al. 2004, and Willmer et al. 2006 for a description  of the DEEP2 survey, catalog construction, 
and data reduction.
Unlike the other DEEP2 fields, no color pre-selection was applied,  therefore
$z < 0.7$ objects were {\it not} excluded as spectroscopic targets. 
About seventy percent of objects brighter than $R (AB) =24.1$ were targeted by DEEP2, 
with $\sim$ 80\% of those yielding 95\% confidence (quality $\ge 3$) redshifts, for a 
total spectroscopic redshift completeness $\sim$ 56\% (Willmer et al. 2006).
We have 3839 matches of quality $\ge 3$ DEEP2 redshifts to the ACS photometry catalog,
and 3000 matches of quality $\ge 3$ DEEP2 redshifts to ACS galaxies with good morphologies. 
Because we do not require high precision redshifts to examine the morphological
evolution with redshift, we supplemented our spectroscopic redshift catalog with 
photometric redshifts determined by Ilbert et al. (2006) using the CFHTLS in $ugriz$. 
The CFHTLS Deep Field D3 surveys the EGS region down to a limiting $i$(AB) $\sim$ 26.2, 
but has small gaps in its coverage due to gaps in the MegaCam CCDs. We use only photometric redshifts
calculated with 3 or more CFHTLS bands including the $i$ band.  We have 52548 matches of 
CFHTLS photometric redshifts with more than 2 bands to the ACS photometry catalog, and 
10140 matches to ACS galaxies with good morphologies.   5960 CFHTLS galaxies with
more than 2 bands have been matched to DEEP2 quality $\ge$ 3 redshifts; 2634 of these have
good ACS morphologies.
The CFHTLS photometric redshifts from Ilbert et al. (2006) are reliable to
$|\Delta z| \leq 0.07 (1+z_{spec})$ for $z_{spec} \le 1.2$, $i > 24.0$, and $g-i > 0.5$, with typical
$\sigma(\Delta z) = 0.04 (1+z_{spec})$ (Figure 1). The catastrophic failure rate where $\Delta z > 0.1 (1+z_{spec})$ 
when compared to the DEEP2 quality $\ge$ 3 redshifts is less than 4\% at $0.2 < z_{phot} < 1$.  This
rate increases to $\sim$ 8\% for $z_{phot} > 1$ with typical $z_{spec} \sim 0.7$.  Catastrophic
photometric redshift errors have a negligible effect on our morphological fractions (\S 3.2).
We performed bootstrapped simulations of
the photometric redshifts and  morphological distributions as a function of redshift (\S 3.2) and find that the
photometric redshift errors have no effect on observed morphological evolution.

Although the DEEP2 and CFHTLS redshift catalogs probe out to $z\sim 1.5$, we select objects 
at $z \le 1.2$ where the ACS images sample rest-frame optical ($\lambda > 3700$\AA\ ) morphologies
to minimize rest-frame wavelength-dependent morphology biases. 
Galaxies appear significantly more irregular in rest-frame ultra-violet than in rest-frame
optical images, and strong biases in the fraction of disturbed objects can occur when
the wavelength dependence on morphology is not taken into account (e.g. Hibbard \& Vacca 1997, 
Marcum et al. 2001,  Windhorst et al. 2002).
In order to consistently sample rest-frame $B$ ($\sim$ 4500 \AA) morphologies across the full redshift 
range, we use the observed $V_{F606W}$ morphologies at $0.2 < z < 0.6$ and the $I_{F814W}$ 
morphologies at $0.6 < z < 1.2$ for our morphological classifications (\S 3). 
The change in morphology between rest-frame 5100\AA\ (for $V$ at $z \sim 0.2$ and $I$ 
at $z \sim 0.6$) and rest-frame 3700\AA\ (for $V$ at $z \sim 0.6$ and $I$ at $z \sim 1.2$)
is small for most local galaxies (LPM04). However we do note that the $z \sim 0.6$ and $z \sim 1.2$
sample have slightly higher merger fractions, possibly the result of this shift in rest-frame wavelength.

\subsection{Completeness Issues}
Galaxies in our final morphological sample must have (1) reliable rest-frame optical morphological measurements, 
i.e. $\langle S/N \rangle$ per pixel $\ge 2.5$, $r_p \ge 0.3 
\arcsec\ $, and contiguous segmentation maps and
(2) a reliable spectroscopic or photometric redshift, i.e. quality $\ge 3$ for DEEP2 $z_{spec}$ or
3 or more bands used for CFHTLS $z_{phot}$.  Therefore the completeness of our matched morphology-redshift
catalog depends on the detection limits in the ACS images, the surface-brightness and size limits of 
our morphological measurements, and the completeness of the DEEP2 and CFHTLS redshift catalogs which may 
be a function of observed color as well as magnitude.  The detection limits of the ACS data are substantially
deeper than our morphology and redshift catalogs, thus we use the ACS photometric catalog to 
compute completeness as a function of SExtractor 'BEST' magnitude, $r_p$, 
and $V-I$ color for the morphology and redshift catalogs (Figures 2 and 3).  Stars are excluded from the
ACS photometric catalog by removing objects with SExtractor stellarity class $> 0.9$.

The completeness of the combined morphology and redshift catalogs as a function of Petrosian radius and
magnitude are shown in $V$ and $I$ in the right-hand panels of Figure 2. 
The Gini coefficient and $M_{20}$ become unreliable at $\langle S/N \rangle$ per pixel $\la$ 2.5 (LPM04, Lotz et al. 2006).
This $\langle S/N \rangle$ per pixel limit corresponds to average surface brightness within the Petrosian
radius $\mu \sim$ 24.7 AB magnitudes per arcsec$^2$ in $V_{F606W}$ and $\sim$24.2 in $I_{F814W}$ for face-on
galaxies (solid line, right-hand panels in Figure 2).  Brighter than these
surface brightness limits, the joint morphology-redshift catalog is $>$ 70\% complete when compared to the ACS photometric catalog.
Compact distant galaxies are also excluded because it becomes impossible to recover $G$ and $M_{20}$ 
for galaxies with $r_p$ less than 0.3 $\arcsec$ in ACS images (Lotz et al. 2006). However, these are less than
6\% of the $I<25$ sample and are mostly at $I > 24$ (Figures 2 and 4).   The DEEP2 spectroscopic redshift targeting selection 
excludes objects with surface brightnesses fainter than $\mu_R \sim 26.5$ (Davis et al. 2007), 
while the CFHTLS photometric redshift 
catalog has no strong selection against low surface brightness objects when compared to the ACS detections (Figure 2).   
Therefore the dominant selection effect is the surface-brightness limit of the morphology catalog. 

One may expect the redshift completeness to be a function of observed color for 
both the spectroscopic and photometric redshift catalogs.  DEEP2 spectroscopic redshifts are 
determined using either strong emission lines found in blue star-forming galaxies 
(e.g. [OII] 3727, H$\beta$, [OII]5007, H$\alpha$) or absorption lines in older red stellar populations, 
while photometric redshifts are most reliable for older and redder stellar population with strong spectral
breaks (e.g. 4000\AA\ break).   However we do not find a strong net bias ($\delta$ completeness $>$10\%) against galaxies 
with either red or blue observed $V-I$ colors in our final morphology-redshift catalog (last panel, Figure 3).
The DEEP2 spectroscopic catalog is  more likely to miss faint ($I> 21$) red ($V-I > 1.0$) galaxies;
however, the CFHTLS photometric redshift catalog is $>80\%$ complete at $21 < I < 24$, $V-I > 1.0$. Hence
the final combined catalog including both spectroscopic and photometric redshifts 
is not strongly biased against red galaxies.   In fact, we are slightly 
biased against blue galaxies (lower left hand panel of Figure 3).  This is because the lowest surface brightness
galaxies missing from the morphology catalog are more likely to be blue.   We
will discuss the implications for the observed morphology evolution in the next section.  
For objects with surface brightnesses greater than the 70\% completeness limit found
in Figure 2 (dashed line), we do not find a significant dependence of completeness on observed 
color (last panel, Figure 3).

\begin{figure*}
\epsscale{1.0}
\plotone{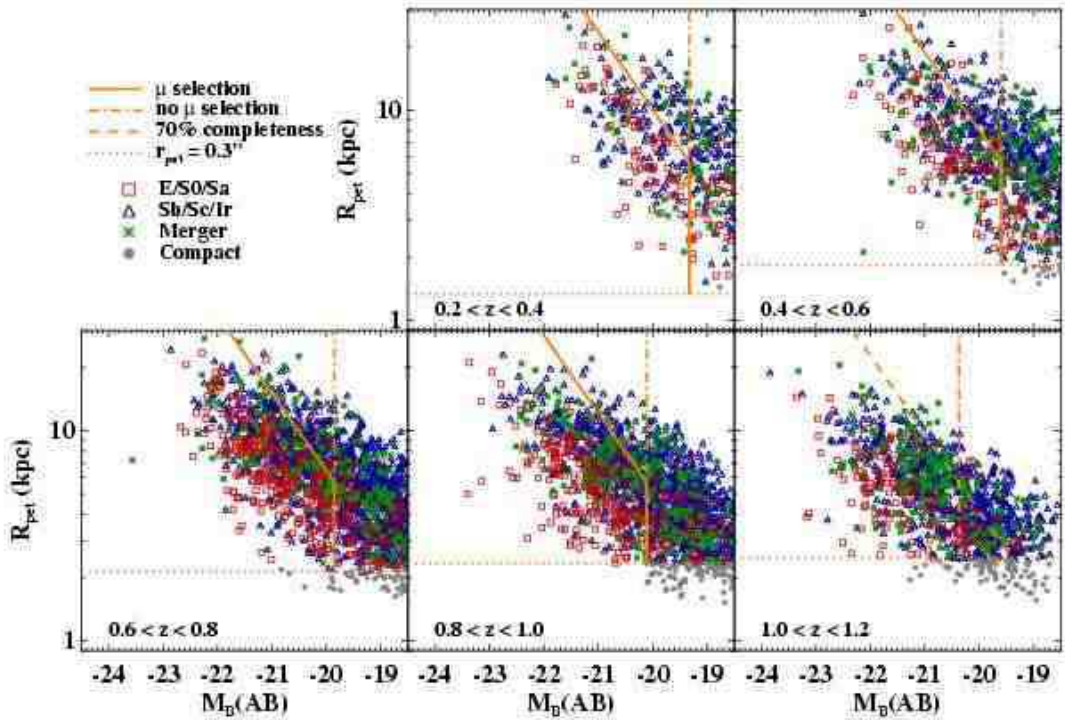}
\caption{Sample selection boxes in intrinsic size and rest-frame magnitude v. redshift. 
We plot Petrosian radius (kpc) v. $M_B$(AB) for all objects in our joint morphology-redshift catalog
with good morphologies and redshifts.  The objects classified by $G-M_{20}$ as E/S0/Sa are shown as 
red squares, Sb-Ir as blue triangles, and mergers as green asterisks.
Objects with $R_{pet} < 0.3 \arcsec\ $ (orange dotted lines) are too compact to classified and are shown
as grey circles. The dashed orange line in the $1.0 < z < 1.2$ panel corresponds to the 70\% completeness
limit shown in Figure 2, while the dot-dashed orange line in this panel corresponds to $I \sim 24$. 
We assume no size evolution and pure luminosity evolution of 1.3 $M_B$ per unit redshift to select 
our volume-limited luminosity evolution sample.   Applying the intrinsic size-magnitude cut where the highest 
redshift bin is $>$70\% complete gives the solid orange selection boxes in the lower redshift bins.  
If we instead apply only a luminosity selection, we use the vertical cuts given by the dot-dashed orange lines.   
The $>$ 70\% completeness cut primarily excludes Sb-Irs.}
\end{figure*}

\subsection{Volume-limited sample selection}
To track the evolution of galaxy morphology with redshift, we must first define a galaxy sample 
that is statistically complete in morphology and redshift measurements and  probes the same
population of galaxies (in terms of mass or luminosity) across the entire redshift range.
In the previous sections, we found that the EGS matched morphology-redshift catalog is $>$ 70\% complete
for objects with $I \leq 24$ and $\mu_{I} \leq 24.2$ with no color biases within this surface
brightness limit.  At $z = 1.2$, $I=24$ corresponds to an absolute magnitude limit of $M_B$ (AB) $\sim -20.5$
(approximately 0.4 $L_B^*$, adopting $M^*_B$ (AB) = -21.44 from Faber et al. 2007).   
Absolute $M_B$ and rest-frame $U-B$ colors are computed from the DEEP $BRI$ photometry with empirical
SED templates  and from the ACS $VI$ photometry where the DEEP $BRI$ 
photometry is not available (see Willmer et al. 2006 for k-correction methodology).

We have constructed a volume-limited sample to study the morphology evolution 
from $0.2 < z < 1.2$ by applying the $>$70\% completeness size-magnitude cut 
to our highest redshift bin ($1.0 < z < 1.2$) and making certain assumptions about how
these galaxies evolve in luminosity and size with time (Figure 4).  
We assume no size evolution.  We assume pure luminosity evolution with an increase
of 1.3 $M_B$ per unit redshift for all galaxies (Faber et al. 2007), and 
apply a cut of $L_B > 0.4 L^{*}_{B}$.   These
limits correspond to $M_B <= -18.94 - 1.3z$, and $M_B < -2.70$ log10(r$_p$) $- 16.90 - 1.3z$ (Figure 4).
Our final volume-limited sample has 2565 galaxies.   We find that our results are not strongly dependent 
on the magnitude of the assumed luminosity evolution, which
may be between 1.0 and 1.5 magnitudes (e.g. Barden et al. 2005, Brown et al 2006, Faber et al. 2007, 
Melbourne et al. 2007).   Roughly 60\% of this sample  has
DEEP2 spectroscopic redshifts, and we use CFHTLS photometric redshifts for the remaining 40\% of the sample.
The redshift distributions of our luminosity evolution $L_B > 0.4 L^*_B$ sample with the 70\% completeness
size-magnitude cut are shown in Figure 5.  The spectroscopic and photometric redshift sub-samples
show similar redshift distributions. The photometric redshift distributions do not detect over-densities
at $z \sim 0.72$ and $z \sim 0.85$ found by the DEEP2 survey and contribute $\sim$ 50\% of the sample
at $z > 1$. 
If we do not apply the $>$ 70\% completeness size-magnitude cut, we have a sample of 3009 galaxies.

The disadvantage of the size-magnitude completeness cut is that we exclude many objects with perfectly good redshifts 
and morphologies with lower surface-brightnesses from the lowest redshift bins (Figures 4 and 5).   
We stress that the conclusions of this paper apply only within the sample selection criteria 
-- i.e. for galaxies with relatively bright optical
magnitudes and high surface-brightnesses. With the EGS data alone, we cannot know how these excluded
objects evolve because we cannot observe the progenitors of these objects in our highest redshift bin if their surface-brightnesses
evolve passively.   We note that this limitation is a problem for all morphological analyses (including visual
classifications) which properly take into account the surface-brightness selection effects and biases. 
Most of the excluded objects are classified as late-type spirals and have blue colors (Figures 3 and 4); 
late-types have larger Petrosian radii and lower-surface brightnesses than early-types at a given magnitude.  Some excluded objects
are mergers, but we find mergers are just as likely to have high surface brightnesses as
low-surface brightnesses and so are not preferentially removed from our samples in the lower redshift bins.   
Therefore the $>$ 70\% completeness cut effectively lowers the ratio
of late-types to early-types.  A preliminary morphological analysis of the Great Observatories Origins Survey (GOODS) 
ACS data (Giavalisco et al. 2004), which probes
lower-surface brightnesses but a smaller volume at $z \sim 1.1$, supports this conjecture. 
We calculate the morphological fractions for a $L_B > 0.4 L^*_B$ sample that includes all low surface brightness
objects in \S 3.2 (dot-dashed lines shown in Figure 4) and compare to the $>$ 70\% complete volume-limited sample described above. 

For each object we determine the completeness for its magnitude, size, and $V-I$ color using
the contours in Figures 2 and 3. The average completeness for each redshift bin and morphological type are
given in Table 1. We also compute a correction factor for the morphological fraction W(type) $=$ $\langle$C(total)$\rangle /
\langle$C(type)$\rangle$.  These corrections are quite small, and never result in changes of more than 2\%.  The
fractions given in Table 2 are corrected by the values given in Table 1.

\begin{figure}
\epsscale{1.0}
\plotone{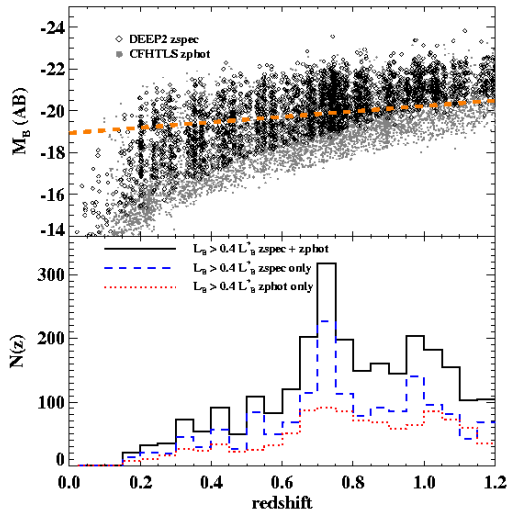}
\caption{{\it Top:} Absolute magnitude $M_B$ v. redshift for all $I < 25$ galaxies with good morphologies and redshifts.  
The orange dashed line shows the $L_B > 0.4 L_B^*$ cut . 
{\it Bottom:} The redshift distribution of the volume-limited $L_B > 0.4 L_B^*$ sample  
with the 70\% completeness cut (black solid histogram), 
and the spectroscopic redshift (blue dashed histogram) and photometric redshift (red dotted
histogram) sub-samples.}
\end{figure}

\begin{figure*}
\plotone{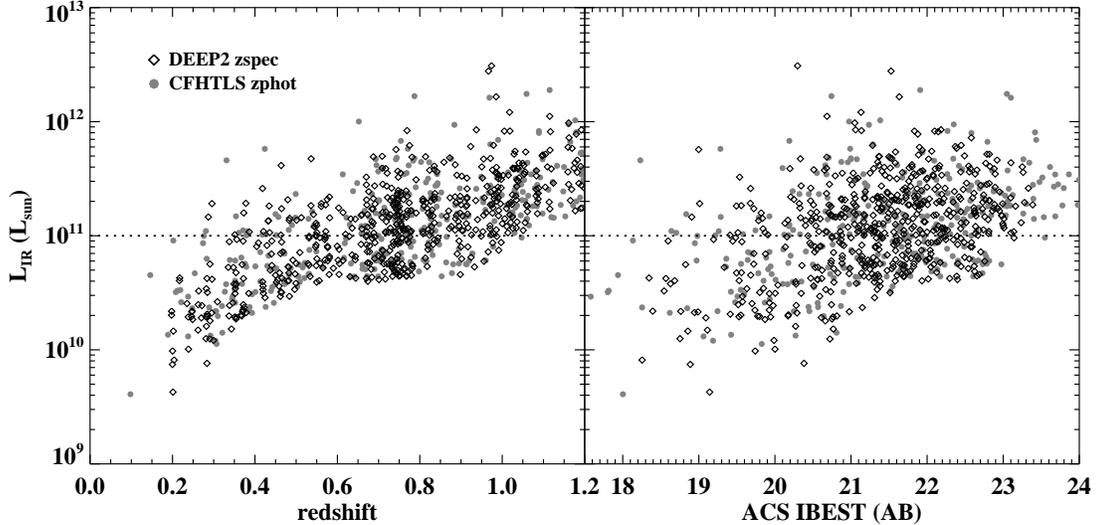}
\caption{Infrared luminosity $L_{IR}$ ($L_{\odot}$) v. redshift (left) and $I$ (AB) (right) 
for the MIPS 24$\mu$m sample 
uniquely matched to within 1.5\arcsec\ of an ACS galaxy with good morphology and a redshift.  Luminous infrared
galaxies (LIRGS) lie above the horizontal dashed line at $L_{IR} > 10^{11} L_{\odot}$.
The majority of LIRGS are $\ge$ 1 magnitude brighter than our $I=24$ magnitude cutoff.  }
\end{figure*}

\begin{deluxetable*}{lccccccc} 
\tablecolumns{8} 
\tabletypesize{\footnotesize} 
\tablecaption{Completeness Corrections} 
\tablehead{ \colhead{z} & \colhead{$\langle$C(tot)$\rangle$}  & \colhead{$\langle $C(E-Sa) $\rangle$} & \colhead{$\langle$C(Sb-Ir)$\rangle$} 
& \colhead{$\langle$C(M)$\rangle$}  & \colhead{W(E-Sa)} & \colhead{W(Sb-Ir)} & \colhead{W(M)}} 
\startdata
\cutinhead{$L_B > 0.4 L^*_B$ with completeness cut }
 0.30&	 0.88 &   0.86	& 0.90 &   0.81  & 1.01 & 0.97  & 1.08\\
 0.50&	 0.83 &   0.82	& 0.83 &   0.86  & 1.01  & 1.00 & 0.96\\
 0.70&	 0.85 &   0.87	& 0.84 &   0.85  & 0.98 &  1.01 &  1.00\\
 0.90&	 0.80 &   0.86	& 0.78 &   0.76 &  0.93 & 1.03 &  1.06\\
 1.10&	 0.72 &   0.80	& 0.70 &   0.70 &  0.91 & 1.03 &  1.03\\
\cutinhead{$L_B > 0.4 L^*_B$ with no completeness cut }  
 0.30 & 0.86 & 0.86 & 0.87 & 0.81 & 0.99 & 0.98 & 1.05 \\
 0.50 & 0.78 & 0.81 & 0.78 & 0.84 & 0.97 & 1.00 & 0.93 \\
 0.70 & 0.81 & 0.86 & 0.79 & 0.81 & 0.94 & 1.02 & 1.00  \\
 0.90 & 0.76 & 0.85 & 0.74 & 0.73 & 0.90 & 1.03 & 1.05\\
 1.10 & 0.70 & 0.79 & 0.68 & 0.67 & 0.88 & 1.02 & 1.04\\
\cutinhead{$L_{IR} > 10^{11} L_{\odot}$}
 0.30 & 0.91 & 1.00 & 0.97 & 0.80 & 0.91 & 0.94 & 1.14 \\
 0.50 & 0.82 & 0.75 & 0.79 & 0.92 & 1.09 & 1.04 & 0.90 \\ 
 0.70 & 0.82 & 0.82 & 0.83 & 0.84 & 1.00 & 0.99 & 0.98 \\
 0.90 & 0.79 & 0.86 & 0.78 & 0.79 & 0.92 & 1.01 & 1.00 \\
 1.10 & 0.73 & 0.81 & 0.73 & 0.73 & 0.90 & 1.00 & 1.00\\
\enddata
\tablecomments{ The mean completeness for each redshift bin and
morphological type and the correction factors to the morphological
fractions W(type)$=$ C(tot)/C(type) are given. }
\end{deluxetable*}

\subsection{MIPS 24 $\mu$m sample}
{\it Spitzer Space Telescope} Multiband Imaging Photometer for Spitzer (MIPS) data in the 24 $\mu$m band was obtained for the EGS as part of
the GTO program (PI Rieke, Fazio).  We matched the ACS detections to the MIPS 24 $\mu$m catalog (see Papovich et al. 2004
for details of the catalog construction), and found 2902 unique matches brighter than 60 $\mu$Jy and within 1.5\arcsec\ of an ACS detection. 
These were then matched to the redshift catalogs: 889 have spectroscopic redshifts with quality $\ge$ 3, 
while an additional 898 have photometric redshifts.  Infrared luminosities (8-1000$\mu$m, Sanders \& Mirabel 1996)
were derived using the Chary \& Elbaz (2001) templates in the same manner as Le Floc'h et al. (2005). 
Our final LIRG sample of 515 objects was selected to have $0.2 < z < 1.2$,  L(IR)$ \ge 10^{11} L_{\odot}$, 
$\langle S/N \rangle$ per pixel $\ge 2.5$, $r_p > 0.3$\arcsec, and meet the same $>$ 70\% completeness criteria shown
in Figure 4.  Above $z \sim 1$, the 24$\mu$m sample is incomplete at L(IR) $< 2 \times 10^{11} L_{\odot}$ (Figure 6).
Most of the LIRGs in our sample are $\ge$ 1 magnitude brighter than our $I=24$ cutoff magnitude.

\section{Morphology Evolution at $0.2 <  z < 1.2$}

\begin{figure*}
\plotone{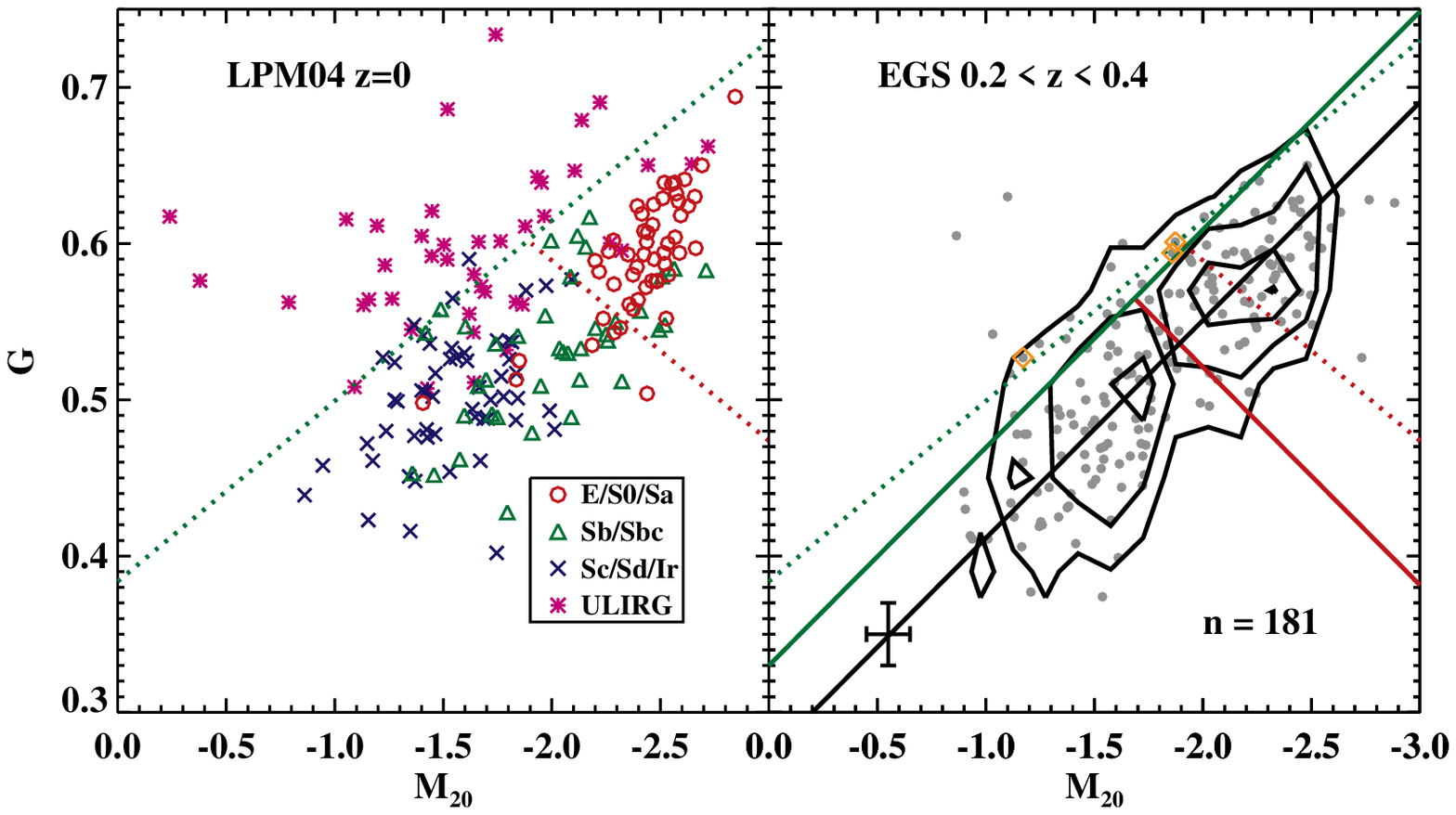}
\caption{{\it Left:} Rest-frame $B$ $G-M_{20}$ morphologies for local galaxies 
from LPM04.  The upper green dotted line divides merger candidates
(ULIRGS) from normal Hubble types (see LPM04). 
The lower red dotted line divides normal early-types (E-Sa) from late-types (Sb-Ir).  
{\it Right:}  Rest-frame $B$ $G-M_{20}$ morphologies for  EGS $0.2 < z < 0.4$ sample (grey points and contours). 
This statistically complete sample has a well defined correlation in $G-M_{20}$ (black line) and shows evidence 
for bi-modality between normal early-types and late-types (contours).  Objects above the solid green line have $G$ $>$ 
3 $\sigma$ above the main locus in $G-M_{20}$ and overlap with the region where $z=0$ mergers lie (dotted green line).
Galaxies above this line but with no visual sign of interactions are marked with orange diamonds.   
Objects above the solid red line occupy the early-type peak in the
observed bi-modality and overlap with the region where $z=0$ E/S0/Sas lie (dotted red line).} 
\end{figure*}

\begin{figure*}
\plotone{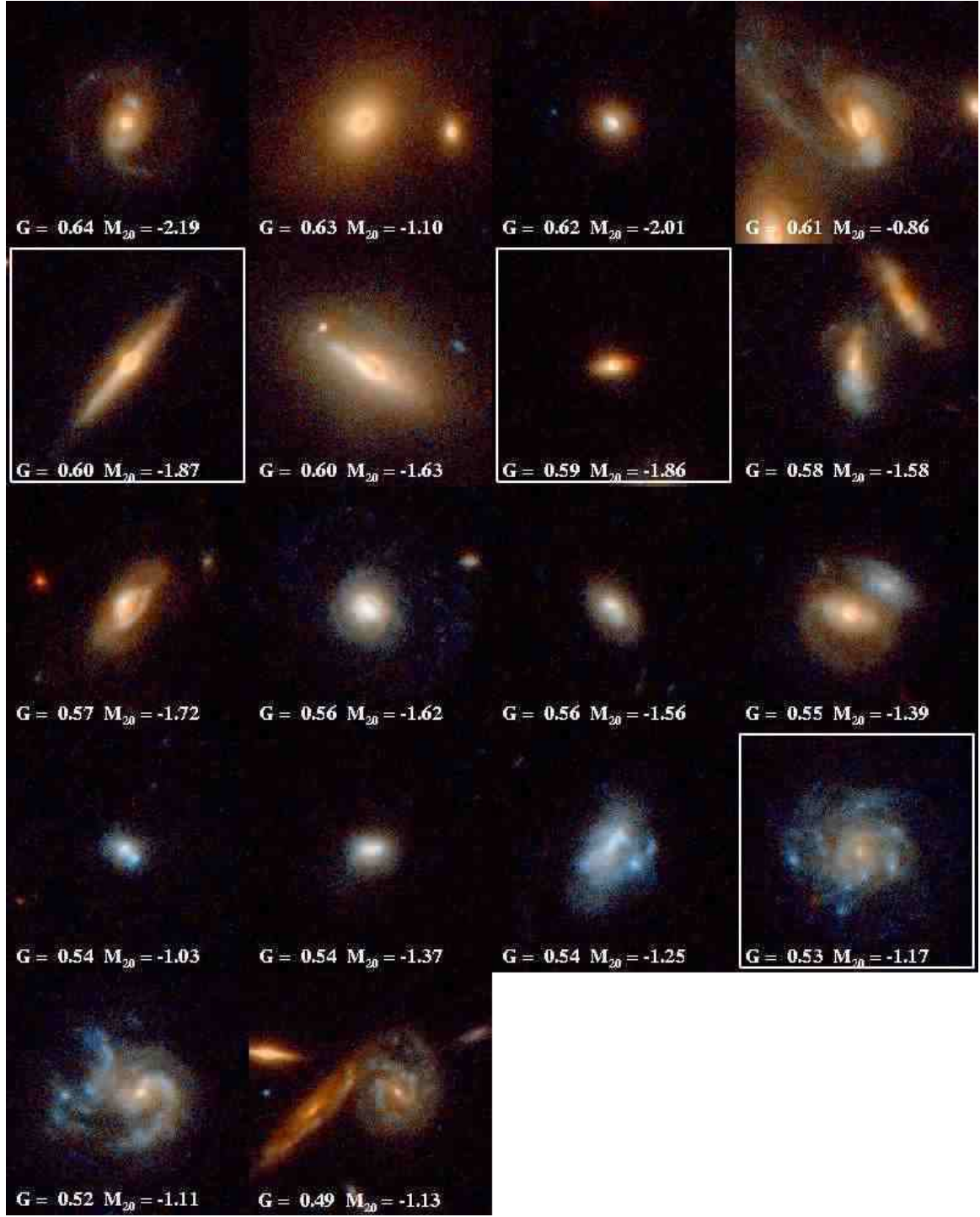}
\caption{ACS $V+I$ color images of merger candidates from EGS $0.2 < z < 0.4$.  These 18 galaxies lie above the solid magenta line in
the right-hand side of Figure 7.  Three of these merger candidates show no signs of recent interaction (boxed images); 
one is an edge-on disk, one is an early-type E/S0, and one is a face-on disk with knotty star-formation. The images are $7.5\arcsec \times 7.5\arcsec$.}
\end{figure*}

\begin{figure*}
\plotone{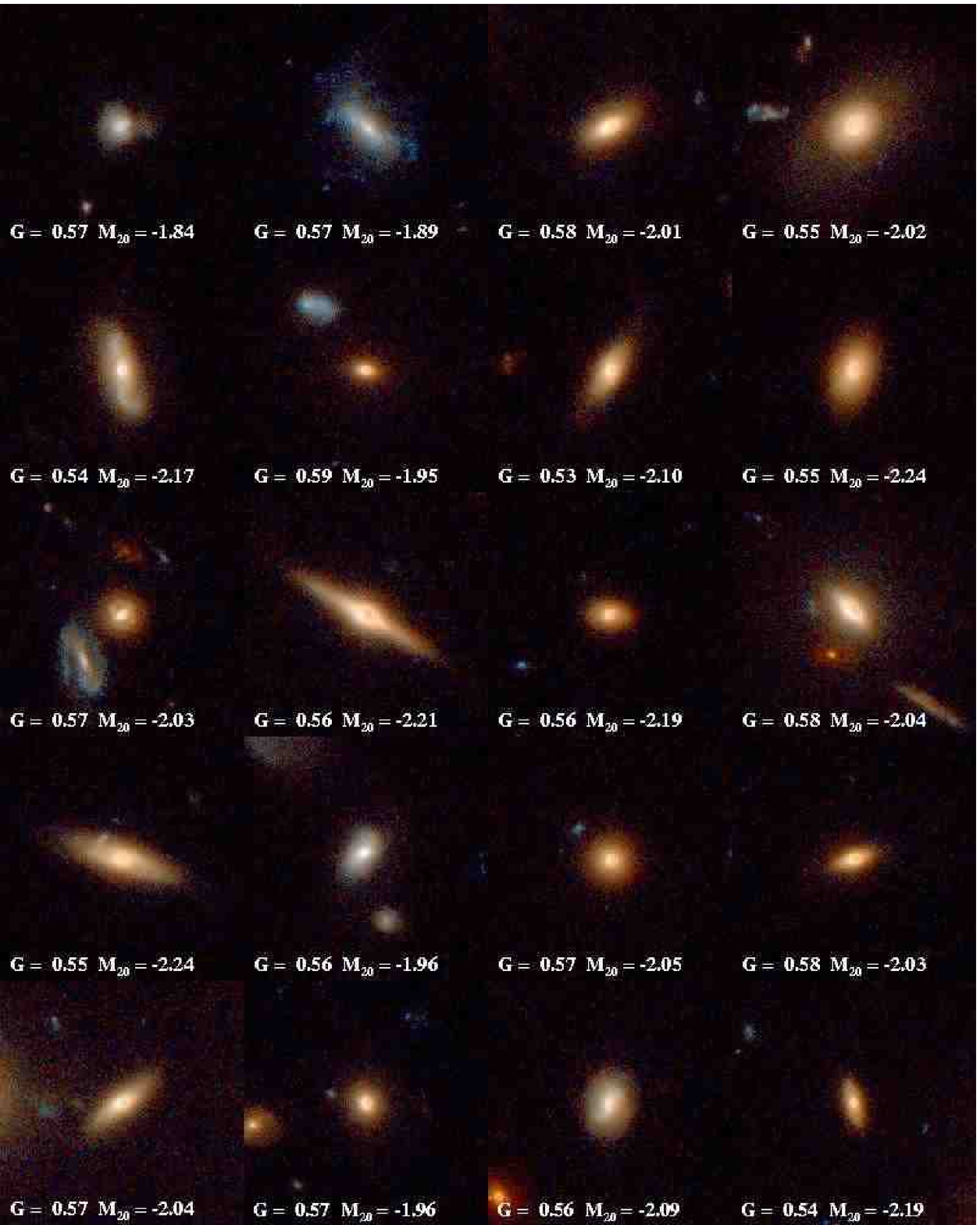}
\caption{ACS $V+I$ color images of border-line early/late type galaxies in the EGS at $0.2 < z < 0.4$.  These galaxies lie between the
solid and dotted red lines in Figure 7.   The majority are bulge-dominated galaxies with little evidence for spiral
arms.  We classify these as E/S0/Sa and adopt the solid red line in Figure 7 
as the division between E/S0/Sa and Sb-Ir $G-M_{20}$ galaxy classification. The images are $7.5\arcsec \times 7.5\arcsec$. }
\end{figure*}

\subsection{Morphological classification at high redshifts}
Visual classification of nearby galaxies has a long and distinguished history (see Sandage 2005 for a recent
review).   However, visual classification assigns galaxies into discrete categories while real galaxies 
lie along a continuum of morphological properties.  
Visual classification is inherently subjective and human classifiers of local well-resolved galaxies often disagree
by one or more classes.  Finally, visual classification of distant galaxies become increasingly difficult 
as the spatial resolution and signal-to-noise of the image decreases and mis-classification become more common
(e.g. Brinchmann et al 1998). 

With quantitative morphology measures, one may avoid the issues of subjectivity and can quantify the effects of
low signal-to-noise and resolution on the measured values. But quantitative measures have their own difficulties. 
Quantitative measurements often show a good deal of scatter when compared to visual classifications.   
This is perhaps unsurprising given the uncertainty in visual classifications and 
given that a single quantitative measurement is unable to capture the full richness of  visual classification. 
With  quantitative measurements, one can choose whether to classify an object into discrete categories by comparing to
visual classifications or to characterize the natural distribution of values (e.g. Blanton et al. 2004).  We
explore both approaches to the $G-M_{20}$ classification scheme. 

The Gini coefficient $G$ and the second-order moment of the brightest 20\% of the light, $M_{20}$, are two 
non-parametric measures of galaxy morphology put forth in LPM04.  With these two quantities measured
in rest-frame $B$, it is possible to identify local galaxy merger candidates and classify early- and late-type
galaxies (LPM04).  $G$ is a statistic used frequently in economics to quantify the
distribution of wealth in a population, and was first used to quantify the distribution of light among a galaxy's
pixels by Abraham et al. (2003).  Although generally correlated with concentration for normal galaxies, it is not
sensitive to the location of the brightest pixels and hence shows high values for galaxies with multiple bright
nuclei as well as highly centrally-concentrated spheroidals. 
A efficient way to compute $G$ is to first sort the pixel flux values $f_i$ into increasing order
and calculate
\begin{equation}
G = \frac{1}{|\bar{f}| n (n-1)} \sum^n_i (2i - n -1) |f_i|
\end{equation}
where $n$ is the number of pixels assigned to a galaxy. 
For uniform surface-brightness galaxies, $G$ is zero; and if one pixel has all the flux, 
$G$ is unity (Glasser 1962).

$M_{20}$ is anti-correlated with concentration, with highly concentrated galaxies having low $M_{20}$ values. 
Unlike concentration which is measured in circular apertures, $M_{20}$ directly traces the spatial extent of
the brightest pixels in a galaxy and is more sensitive to merger signatures like double nuclei.  It is also
normalized to the total moment of the galaxy, and thus is less sensitive to inclination.  It is defined as
\begin{eqnarray}
M_{20} \equiv {\rm log10}\left(\frac{\sum_i M_i}{M_{tot}}\right) & {\rm while } 
& \sum_i f_i <  0.2 f_{tot}
\end{eqnarray}
and
\begin{equation} 
M_{tot} = \sum_i^n M_i = \sum_i^n f_i \cdot ((x_i - x_c)^2 + (y_i - y_c)^2)
\end{equation}
where $x_c, y_c$ is the galaxy's center (LPM04). The center is computed by finding $x_c,
 y_c$ such that $M_{tot}$ is minimized.

In the left-hand side of Figure 7, we plot the $G-M_{20}$ values for a heterogeneous sample of local
galaxies measured by LPM04.   The images of normal Hubble types (E-Ir) are from the Frei et al. (1996) catalog
and the SDSS; the $G$-$M_{20}$ values are measured at rest-frame $B$ (Frei galaxies) or $g$ (SDSS) and the visual
classifications come from the Carnegie Atlas (Sandage \& Bedke 1994).  The normal galaxies lie along a well-defined
sequence in $G$-$M_{20}$, with early-types (E/S0/Sa) exhibiting high $G$ and low $M_{20}$ values and late-types (Sb-dI)
exhibiting low to moderate $G$ and higher $M_{20}$ values.   Although there is significant scatter in the $G$-$M_{20}$ values of
a particular visual classification, most local E/S0/Sa lie above and to the right of the dashed red line while most
spiral and late-types lie below and to the left of the dashed red line.  

We have also plotted a sample of ultra-luminous infrared galaxies (ULIRGS) from Borne et al. (2000) with 
$G$-$M_{20}$ values measured by LPM04.  Over 90\% of the ULIRGs show visual characteristics of merger
activity such as multiple nuclei and/or tidal tails (e.g. Cui et al. 2001, Borne et al. 2000), and they span
a range of merger stages from close pairs to ongoing mergers with double nuclei to recently merged 
objects with tidal tails and bright single nuclei.  Detailed kinematic studies of a smaller sample of ULIRGs
indicate that many ULIRGs are mergers of roughly equal mass galaxies (Dasyra et al. 2006).  
We find that the majority of ULIRGs lie above the sequence of normal Hubble types.  In LPM04, we estimated the
upper limit of the local normal galaxy $G$-$M_{20}$ sequence with the dotted green line where major-merger candidates fall 
above this line. 

In the right-hand side of Figure 7, we plot the rest-frame $B$ $G$-$M_{20}$ values for our $>$ 70\% complete
$L_B > 0.4 L_B^*$ sample at $0.2 < z < 0.4$ (see also first panel in Figure 10).  Unlike the local galaxy 
sample studied in LPM04, this sample is $\sim$ 88\% complete (Table 1) and we can look for features in the distribution
of $G$-$M_{20}$ space and use any such features to classify galaxy morphologies. 
We again find the most galaxies lie along a well-defined sequence in $G$-$M_{20}$ (fit by the solid black line). 
The $G-M_{20}$ distribution shows evidence for bi-modality (solid contours), with one peak corresponding to the region 
where local Sb/Sc/Ir fall and the second peak slightly below where local  E/S0/Sa fall. 
The solid red line indicates the minimum between these two peaks. 
The solid green line has the same slope as the line fit to the $G-M_{20}$ locus, but is has a zeropoint shifted
$+0.06$ in $G$ (three times the typical uncertainty in $G$).  Objects above the solid green line 
lie in roughly the same region as the major-merger candidates measured by LPM04 (dashed green line).

We visually inspected the $0.2 < z < 0.4$ sample to determine how well the classification based on the features of the 
$0.2 < z < 0.4$ $G$-$M_{20}$ distribution corresponded with visual morphologies. 
In Figure 8, we show the 18 merger candidates above the solid green line in Figure 7, 
including those objects where the two classification cuts disagreed.
We find that all but 3 objects show morphological merger signatures in the form of very close pairs, off-center nuclei, 
or asymmetries (orange diamonds, right panel of Figure 7).   
One undisturbed object is an edge-on disk, one is an early-type galaxy with an uncorrected
cosmic-ray, and one is a face-on spiral with very knotty star-formation. 
In Figure 9, we show 20 objects that lie between
the solid and dashed red lines in Figure 7.   Almost all of these objects appear to be red S0 or Sa galaxies with a 
significant bulge and a disk component but little evidence for spiral arms.   We conclude that the classification cuts based on the
natural distribution of $G$-$M_{20}$ space do a better job of identifying $z>0.2$ early-type galaxies and finds 
slightly more merger candidates than the $z=0$ visual classification scheme.  We adopt the following classifications for $0.2 < z < 1.2$ EGS galaxies:  
\begin{eqnarray}
\begin{array}{llll}
{\rm Mergers:} &  G > -0.14\ M_{20} + 0.33  &   & \\
{\rm E/S0/Sa:} &  G \le  -0.14\ M_{20} + 0.33  & \&\  G > 0.14\ M_{20} + 0.80  & \\
{\rm Sb-Ir:} &  G  \le  -0.14\ M_{20} + 0.33 & \&\ G \le 0.14\ M_{20} + 0.80 & \\
\end{array}
\end{eqnarray}
The merger  classification cut is shown as the green line in the right panel of Figure 7, the
division between E/S0/Sa and Sb-Ir is shown as the red line in the same figure.  
Changes in the positions of these cuts will change the relative fractions of mergers, E/S0/Sa, and
Sb-Ir.   We present the effects of small offsets to these classifications on the derived
morphological fractions in the next section. While our discrete classifications
of `E/S0/Sa', `Sb-Ir' and `major-merger candidate' adopted here are guided by the visual classifications, strictly speaking
the classifications are labels for particular quantitative regions of $G$-$M_{20}$ space.

\begin{figure*}
\plotone{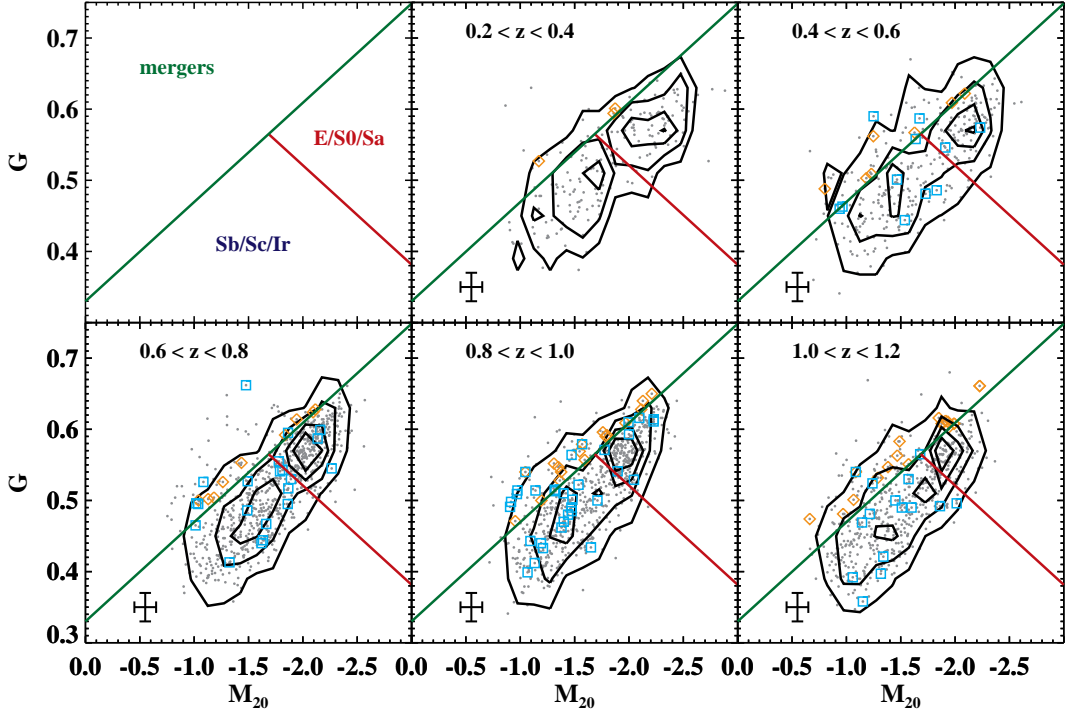}
\caption{$G$ v. $M_{20}$ for the luminosity evolution sample with the $>$ 70\% completeness cut
in EGS as a function of redshift.  The solid lines are the same as right-hand of Figure 7 and
show the division between E/S0/Sa, Sb/Sc/Ir, and merger candidates (first panel).  
Spectroscopically-confirmed chance superpositions are marked as cyan boxes.  Visually
ambiguous merger candidates are marked with orange diamonds. The error-bars show the typical uncertainty for 
$\langle S/N \rangle$ per pixel $=2.5$ galaxies.  The majority of galaxies lie along a well-defined sequence in
$G-M_{20}$, while $\sim$ 10\% lie in the merger candidate region. }
\end{figure*}

The $HST$ ACS observations of the $0.2 < z < 0.4$ sample have worse spatial resolution ($\sim$ 0.5 kpc per PSF FWHM) 
than ground-based images of local galaxies used in LPM04 ($<$ 0.1 kpc per PSF FWHM).  
The shift in $M_{20}$ values from $z=0$ early-types to $z \sim 0.3$ early-types is likely to be caused by
the factor of five change in spatial resolution of the images (see discussion of resolution effects in LPM04).
The change in spatial resolution from $z \sim 0.3$ ($\sim$ 0.5 kpc per PSF FWHM) to $z \sim 1.1$ ($\sim$ 1.0 kpc per 
PSF FWHM) is not as dramatic as the change
from the LPM04 $z = 0$ sample to the EGS $z \sim 0.3$ sample. Nevertheless, we tested
whether the classification cuts based on the $0.2 <$ z $< 0.4$ are robust to redshift-dependent effects.
We artificially redshifted the $0.2 < z < 0.4$ sample to z=0.5, 0.7, 0.9, and 1.1 by rescaling the
galaxy sizes by the angular diameter distance and rescaling the galaxy fluxes by the
luminosity distances and assuming 1.3 magnitudes of brightening per unit redshift. 
We then recomputed $G$-$M_{20}$ values and classifications for the artificially redshifted images.  
We find that the fraction of artificially-redshifted objects that are mis-classified is $\sim$ 15\% and 
does not change significantly with redshift.  Changing the $G-M_{20}$ classification cuts with redshift does not
improve our mis-classification rate. 
The net corrections to the observed morphological fractions are less than 15\% because
for every merger mis-classified as an E/S0/Sa or Sb-Ir, a certain number of  E/S0/Sa or Sb-Ir are mis-classified 
as  mergers. The net corrections are comparable to the Poisson and bootstrapped errors 
derived in the next section and are $-8$\% for Sb-Ir, $+5$\% for E/S0/Sa, and $+$2\% for the 
merger candidate fraction.   Because our artificial-redshift tests do not show a change in the 
mis-classification rate with redshift, 
we adopt the same $G-M_{20}$ classification cuts for all redshift bins.

\subsection{$G$-$M_{20}$ distribution v. redshift}

\begin{figure*}
\plotone{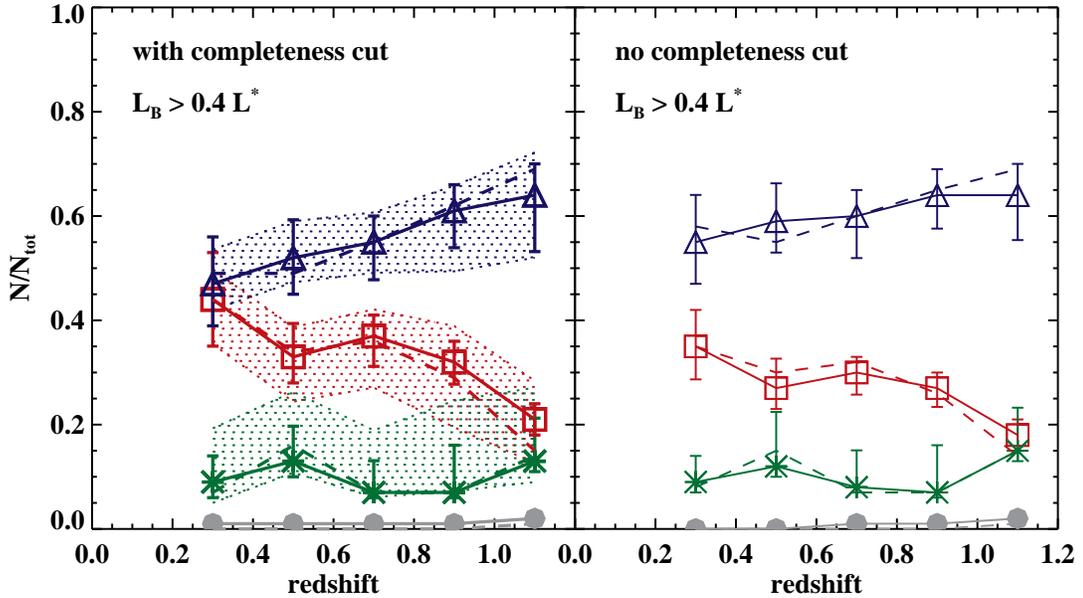}
\caption{EGS morphological fractions as a function of redshift for the luminosity evolution sample with and without
the $>$ 70\% completeness cut.  E/S0/Sa are the
red squares, Sb-Irs are the blue triangles, major merger
candidates are the green asterisks, and compact objects are grey circles. 
Both samples show weak evolution in the fraction of merger candidates and an increase in E/S0/Sa at late times.
The dashed lines give the fractions for objects with spectroscopic redshifts, and the symbols
and solid lines are for the full samples. The shaded regions in the left hand side show
the effect of changing our morphological classifications by $\pm 0.02 G$. }
\end{figure*}

In Figure 10, we plot the $G-M_{20}$ distribution 
of our $> 70\%$ complete volume-limited $L_B > 0.4 L_B^*$ sample as a function of redshift.
The rest-frame $B$-band $G-M_{20}$ distribution at $0.2 < z < 1.2$ is remarkably similar  
to local galaxies, with the majority of galaxies lying along a clear sequence in $G-M_{20}$. 
The error-bars are for a $\langle S/N \rangle$ per pixel $=2.5$ galaxy in ACS images ($\delta$G $=\pm 0.03$,
$\delta$M$_{20} = \pm 0.1$; see Lotz et al. 2006 for tests of $G$ and $M_{20}$ in deep $HST$ ACS images); 
the typical $\langle S/N \rangle$ for our sample is 4-6 with $\delta$G $=\pm 0.025$,
$\delta$M$_{20} = \pm 0.03$.  

We have visually inspected all of the merger candidates, and find that $\sim$ 83\% (332/402) show merger signatures
such as offset or multiple nuclei, visual asymmetries, and close companions.  The remaining $\sim$ 17\% are ambiguous and
usually lie close to the dividing line between mergers and `normal' galaxies (orange diamonds, Figure 10). 
These are often spirals or irregulars with bright off-center star clusters. 
We use the DEEP2 spectroscopic survey to estimate the number of chance
superpositions that result in false merger detections.  Five percent of the galaxies in our sample
with spectroscopic redshifts (83/1530) have multiple systems at different redshifts in a single DEIMOS slit, due to either 
chance superpositions or gravitational lenses (cyan boxes, Fig. 10).  Roughly a quarter (21/83) of the chance
superpositions are classified as merger candidates.  223 of our merger candidates have
spectroscopic redshifts, hence we expect $\sim$ 9\% of all merger candidates to be chance
superpositions.  Extrapolating to the photometric redshift sample, 
we find the correction for chance superpositions has a negligible effect on derived merger fractions
($\sim 1$\%).

Some of our merger candidates are very close pairs with overlapping isophotes and are identified
in the ACS images as a single object by SExtractor.  These objects are also blended in ground-based
images, and the red-blue pairs in particular may have unreliable photometric redshifts.
Sixteen of our merger candidates are very close pairs with only photometric redshifts available.  Nine of
these show $V-I$ color differences $> 0.2$  as measured in 0.5\arcsec\ apertures around
each nucleus in the ACS images.  These blended sources have 
a negligible contribution to the derived merger fraction and evolution.  We have chosen to include them in the
final merger candidate sample, but excluding them would not affect our results.   

We plot the morphological fractions as a function of redshift for the samples with and
without the $>$ 70\% completeness size-magnitude cut (solid lines and points in Fig. 11).  
The fractions have been corrected for incompleteness (Table 1),  
visually ambiguous merger candidates and chance superpositions.  The fraction of objects that
are too compact to classify are plotted as grey points; their contribution is negligible for
all redshift bins.
The solid lines are for the entire sample, and the dashed lines are the spectroscopic
redshift sub-sample.   We find that the fractions derived for objects with spectroscopic redshifts and for the
full samples are very similar, with a slightly lower/higher fraction of early-types/late-types for the spectroscopic
sample in our highest redshift bins. We also find similar evolution in morphologies for the sample with the
size-magnitude cut and the sample with no such cut, with the lowest redshift bins showing
slightly higher fractions of disk-dominated late-type galaxies when no cut is applied. 
The merger fractions are identical. Given the similarity between these two samples, we will only 
refer to the sample with the $>$ 70\% completeness cut for the rest of the paper.

To estimate the uncertainties in the fractions associated with the measurement errors, 
we have also created 10,000 bootstrapped realizations of 
the $G$-$M_{20}$ distributions and compute the average morphological fractions
and 1-$\sigma$ standard deviations for each redshift bin. 
These simulations resample the $G$ and $M_{20}$ measurements using 
uncertainties and biases as a function of $\langle S/N \rangle$ found in Figure 2 of Lotz et al. (2006)
and assuming a Gaussian error distribution.
The simulations also resample the photometric redshifts assuming a photometric redshift
$\sigma = 0.07 (1 + z_{phot})$.  
In general, we derive similar morphological distributions from both the observed samples
and bootstrap realizations, although the simulations classify slightly more galaxies as mergers
and fewer as late-types. 
The error-bars plotted for the observed fractions in Figure 11 are the Poisson uncertainties and 
mean bootstrap shifts added in quadrature. The observed morphological fractions,  
Poisson uncertainties, and mean bootstrap shifts are given in Table 2. 
Our error-bars do not include cosmic variance. 
When we make a jackknife estimate by dividing
the EGS field into 7 regions, we find the standard deviations of the morphological fractions are less than 
expected from Poisson statistics, implying that the effects of cosmic variance are small. 

We show the effect of $\pm 0.02 G$ shifts in the classification criteria given in Eqn. 4. 
The shaded regions show the maximum and minimum values of the morphological fractions when these
shifts are applied.   Although the absolute values of the fractions are sensitive to the
positions of the classification cuts, the evolution in the fractions is unchanged.   
Shifting the merger criteria by $-0.02 G$ roughly doubles the number of objects classified as mergers;
this is because the merger cut is parallel to the main locus of normal galaxies.  This cut includes many
more normal galaxy interlopers, hence we adopt the more conservative merger definition given in Eqn. 4. 

We find that the fraction of major merger candidates does not strongly evolve with redshift at $0.2 < z < 1.2$
for both samples.  The merger fraction is roughly constant at $10 \pm 2$\%  for $0.2 < z < 1.2$
(Figure 11 and Table 2). The observed fraction of Sb-Ir declines from $64 \pm 0.05$\% at $z \sim 1.1$ to 
$47 \pm 0.09$ \% at $z \sim 0.3$ (Fig. 11, blue triangles).  
Many of these systems are ``irregular'' in visual morphology, but have $G$ and $M_{20}$ consistent with 
late-type disks. The observed fraction of E/S0/Sa increases from $21 \pm 5$\% at $z \sim 1.1$ to  $44 \pm 0.09$
at $z \sim 0.3$ (Fig. 11,  red squares).  

The most important sources of possible bias in our merger fractions are incompleteness and classification
errors.  The $z \sim 1.1$ redshift bin is the least complete and if many $z \sim 1.1$ mergers are
low surface brightness, we could underestimate the merger fraction and the evolution in the
merger rate.  However, given the distribution of sizes and magnitudes of the $z \sim 1.1$ merger candidates
we do observe, this is highly unlikely.  The completeness corrections computed in Table 1 strongly
suggest that we are not preferentially biased against high redshift mergers.   Classification
errors are also likely to be significant at high redshift, and our artificial redshift tests
imply that up to 15\% of our sample will be misclassified.  However, this misclassification rate
does not change significantly between $z \sim 0.5$ and $z \sim 1.1$, and the net correction to the
merger fraction is only $\sim$ 2\%.   The weak dependence of the derived morphologies on redshift
shows that redshift errors have little effect on our results so long as they are confined to a
minority of the sample.

\begin{figure}
\plotone{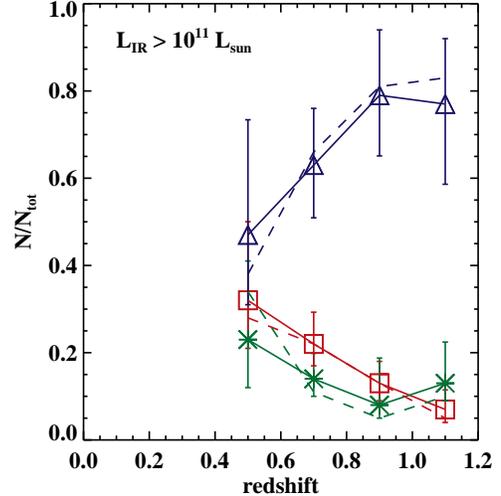}
\caption{LIRG morphological fractions as a function of redshift. The symbols and lines are the
same as the previous plot.  Only 3  LIRGs were too compact to classify and are not plotted here. 
The majority of high redshift LIRGs are Sb-Ir galaxies (blue triangles).}
\end{figure}

\subsection{ Morphologies of LIRGS}

Infrared-luminous galaxies with L(IR) $> 10^{11}$ L$_{\odot}$ are dominated by late-types 
(45-80\%) at $0.4 < z < 1.0$ where the 24 $\mu$m sample is most complete (Table 2; Fig. 12).
Merger candidates are $\sim$ $15\pm2$\% of the LIRG 
sample over the same redshift range (Table 2; Fig. 12).  There is some indication
that the merger fraction of LIRGs may increase to $>$ 25\% at $z < 0.5$, 
although this increase is not statistically significant because of our small sample size at
these redshifts.  
We also find a number of bright 24 $\mu$m sources with E/S0/Sa morphologies (8-30\%).   
Many of these have an obvious star-forming disk (i.e., Sa).  One spheroid-dominated galaxy
has L(IR) $\sim 3 \times 10^{12} L_{\odot}$ and appears to be a post-merger remnant.  
Active nuclei (which may be more prevalent in bulge-dominated systems) may 
be present in $\sim$ 15\% of $z \sim 1$ MIPS sources (Le Floc'h et al. 2005, Bell et al. 2005).

\begin{figure*}
\epsscale{1.0}
\plotone{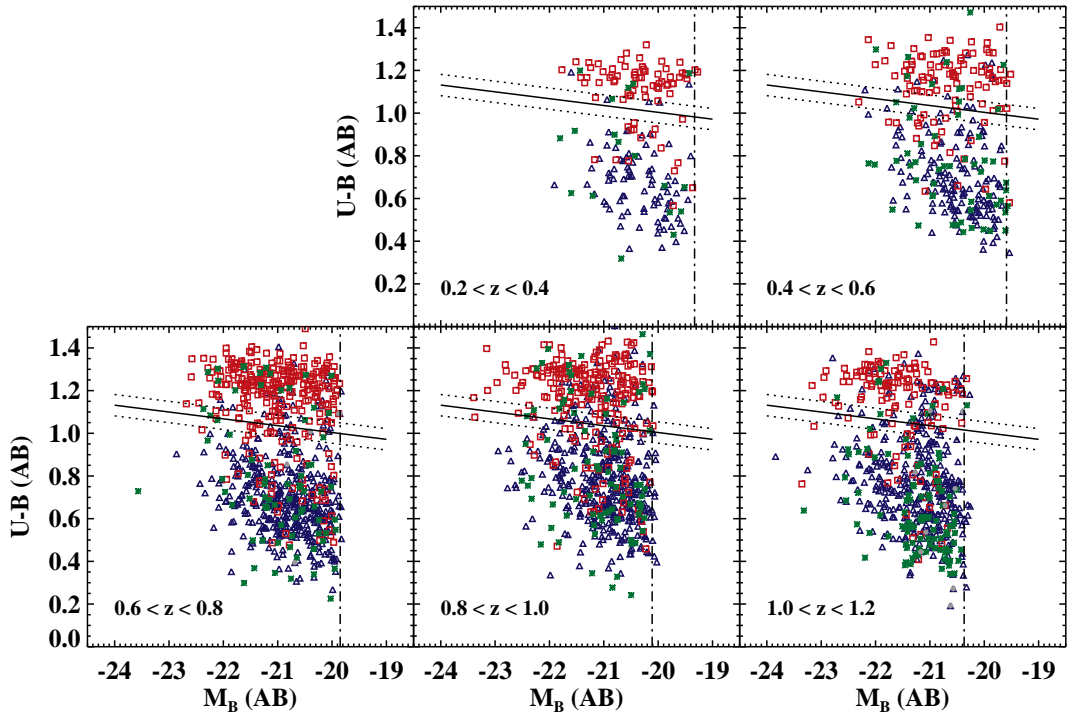}
\caption{$U-B$ v. $M_B$ (AB) for the EGS sample. The symbols indicate the morphological type
(same as previous figures). The vertical dot-dashed line shows the
evolving $L_B = 0.4 L_B^*$, $M_B^* \propto 1.3 z$ cut. The solid line 
is the dividing line between the red and and blue galaxies from Willmer et al. (2006).
Galaxies above the upper dotted line are red sequence galaxies, galaxies below the lower
dotted line are blue cloud galaxies, and galaxies between the dotted lines are
green valley galaxies. Morphological classification correlates well 
with rest-frame color, with $\sim$ 76\% of E/S0/Sa lying on
the red sequence and $\sim$ 86\% of Sb-Ir in the blue cloud.}
\end{figure*}

\begin{figure*}
\plotone{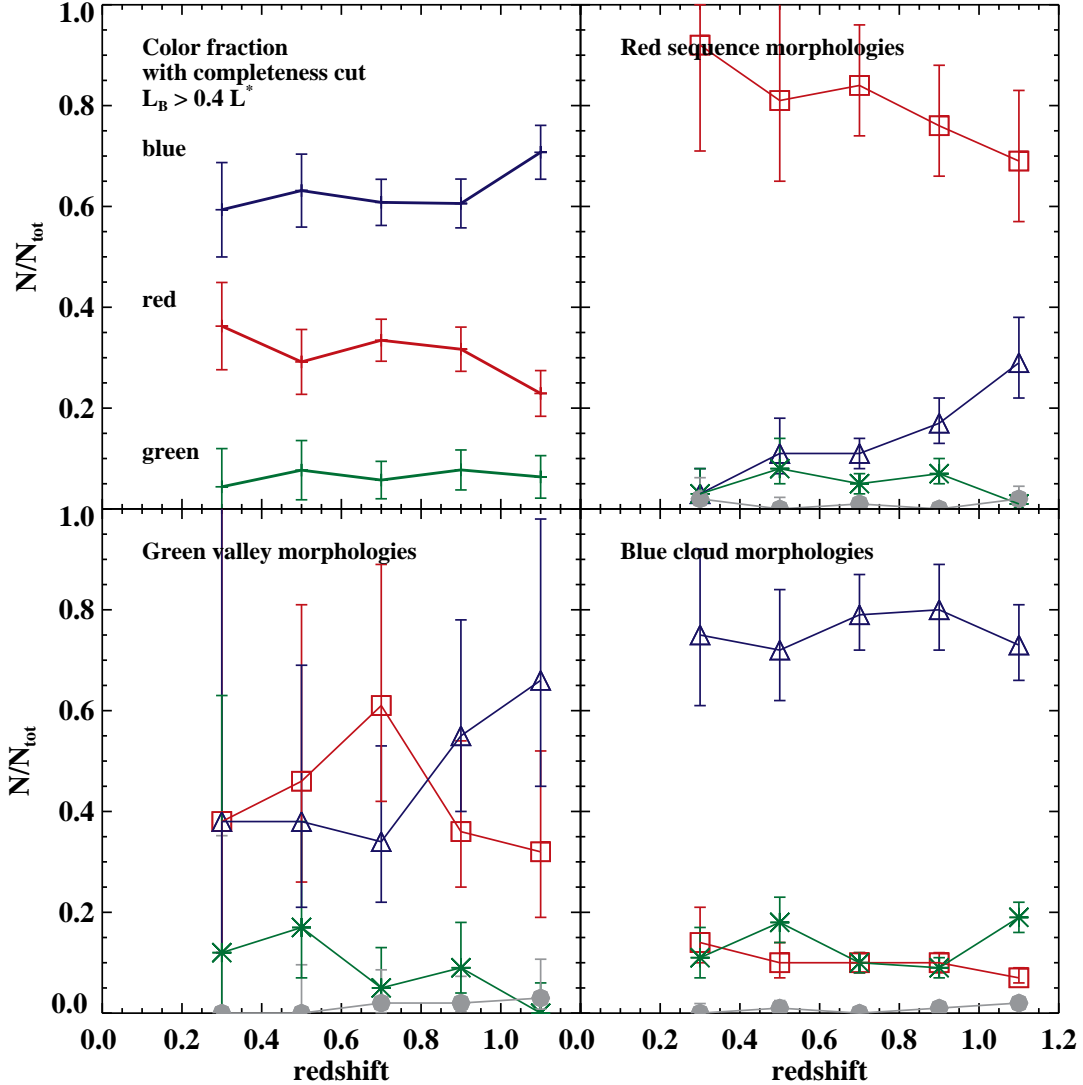}
\caption{Morphological fraction as a function of rest-frame $U-B$ color (see previous figure). 
The symbols are the same as Figure 11.  
{\it Upper left:} The fraction of galaxies in the $>$ 70\% complete $L_B > 0.4 L^*_B$ sample
in the red sequence, green valley, and blue cloud.   {\it Upper right:} The morphologies
of red sequence galaxies v. redshift. {\it Lower left:}  The morphologies of green valley galaxies
v. redshift. {\it Lower right:} The morphologies of blue cloud galaxies v. redshift.}
\end{figure*}

\begin{figure*}
\plotone{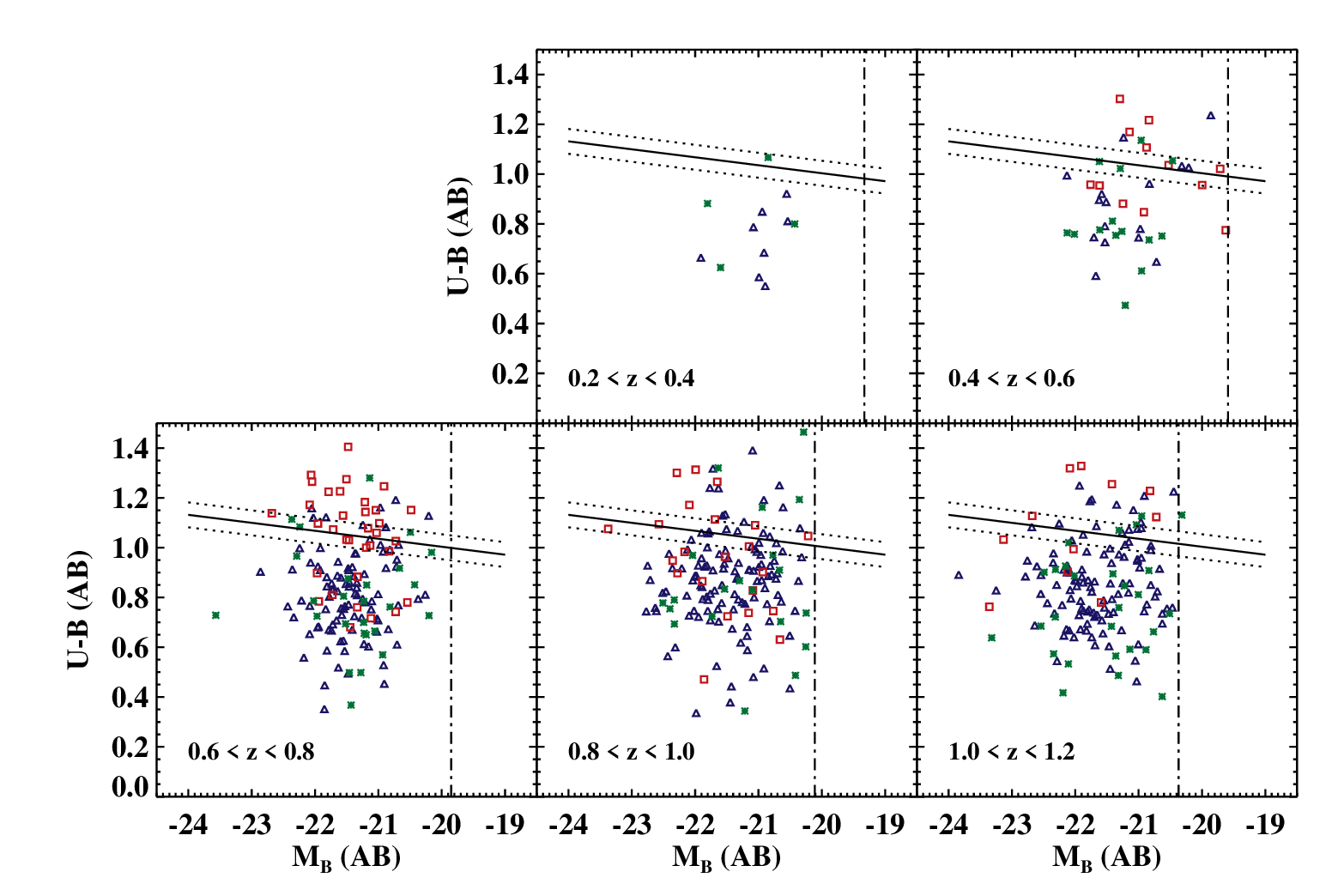}
\caption{$U-B$ v. $M_B$ (AB) for the  $L_{IR} > 10^{11} L_{\odot}$ sample. The majority
of LIRGs are bright blue late-types.  The symbols and
lines are the same as Figure 14. }
\end{figure*}

\section{Morphology-Color Relation and Evolution of the Red Sequence}

The $G-M_{20}$ classifications correlate well with the
galaxies' rest-frame $U-B$ colors (Figure 13). We adopt
the division between red and blue galaxies defined
by Willmer et al. (2006) and define  `red sequence' galaxies
with $U-B$ colors $>$ 0.1 AB mag above this division, 
`blue cloud' galaxies with $U-B$ colors $<$ 0.1 AB mag
below this division, and `green valley' galaxies between
these two cuts :  

\begin{eqnarray}
\begin{array}{lc}
{\rm Red:}   &  U-B  >  -0.032 (M_B + 21.63) + 1.10  \\
{\rm Green:} &  1.00  \le U-B + 0.032 (M_B + 21.63) \le  1.10   \\
{\rm Blue:}  &  U-B <  -0.032 (M_B + 21.63) + 1.00   \\
\end{array}
\end{eqnarray}

Roughly three-quarters of E/S0/Sa lie on the 
red sequence while only 8\% of Sb-Ir do.  Approximately 14\% of mergers 
are red in $U-B$. Conversely, 86\% of Sb-Ir and 80\% of merger candidates 
fall in the blue cloud, while only 18\% of E/S0/Sa are blue.  Given that
60-70\% of our sample are in the blue cloud and 20-40\% are on the red sequence, 
our merger candidates are more likely to be blue than the general population. 

The morphological make-up of the red sequence evolves
with redshift, while the morphologies of blue cloud galaxies remain largely
unchanged. In Figure 14, we have plotted the observed morphology distribution of
the red sequence, green valley, and blue cloud galaxies.  
At $z \sim 1.1$, $69 ^{+14}_{-12}$\% of red
galaxies lie in the E/S0/Sa region of $G-M_{20}$ space, while at $z \sim 0.3$
 $92^{+8}_{-21}$\% of the red sequence galaxies are classified as E/S0/Sa.  
Most of the remaining red galaxies are disk-dominated galaxies,
with Sb-Ir making up $29^{+9}_{-7}$\%  of the red sequence at
$z \sim 1.1$. Some of the red disk galaxies are edge-on systems, and
many are similar to the diffuse red galaxies found by Weiner et al. (2005) in 
DEEP1 WFPC2 images.  The remaining red sequence galaxies
lie in the merger-candidate region of $G-M_{20}$ space.

Blue cloud galaxies are dominated by disk galaxies at all redshifts, and
show a roughly constant fraction of mergers ($\sim$ 13\%) 
and blue E/S0/Sa ($\sim$ 9\%).  The majority of blue E/S0/Sa are Sa with
very bright bulges and blue star-forming disks, although some appear to
be pure spheroids without disks.   There is a suggestion that green valley
galaxies transition from being dominated by disk galaxies at $z \sim 1.1$
to having equal numbers of spheroidal and disk-dominated galaxies at $z < 0.8$, 
but there are too few green galaxies to make a strong conclusion. 

The majority of luminous infrared galaxies are bright ($M_B < -21$) and 
blue ($U-B  \sim 0.65-1.0$; Figure 15).  
However, some are red in rest-frame $U-B$.  Most of these are morphologically classified as 
Sb-Ir,  suggesting that many red disks are reddened by dust. Eight are merger candidates, and may
also be heavily reddened.  Thirty-one of the red LIRGs are E/S0/Sa.  Many of these are Sas with
visible star-forming disks. The remaining red spheroidal LIRGS could host IR-luminous AGN.

\section{Galaxy Major Merger Rate}

The $G-M_{20}$ merger criteria given in Eqn. 4 do not identify all galaxies 
undergoing the merger process.   The sensitivity of $G-M_{20}$ to merger activity
will depend on merger stage, mass ratio, gas content, initial morphologies, initial
orbital conditions, viewing angle, dust extinction, and wavelength.  For example, galaxies during their
first pass may meet the kinematic pair criteria but may not have experienced the
tidal fields necessary to distort their morphologies, and hence would not be classified
as merger candidates using $G-M_{20}$.  On the other hand, recently merged gas-rich galaxies
will exhibit highly disturbed morphologies but are no longer considered a kinematic pair.  
Computation of a {\it merger rate} from the morphological merger fraction
(whether those merger candidates are identified by visual classifications or quantitative
morphologies) requires that the observed merger fraction be weighted by the timescales during
which mergers can be identified via their morphologies.  Those timescales are largely
uncalibrated and will depend on the mass ratios, initial galaxy types, merger orbits, etc.

Our $G-M_{20}$ merger criterion was initially determined using a large sample of ultra-luminous
infrared galaxies (LPM04), which have been shown to be gas-rich mergers with mass ratios $\ge$ 1:3
(Dasyra et al. 2006).  Gas-rich equal-mass merger models suggest the timescales during which merging 
systems lie in the merger candidate region in Figure 10  are 0.5-1 Gyr, depending
on the viewing angle, extinction, and  merger parameters (Lotz et al., in prep).  However, some of our
merger candidates are red spheroids and gas-poor mergers may appear morphologically disturbed for 
shorter timescales ($\sim$ 0.25 Gyr, Bell et al. 2006a).  

The galaxy major merger rate for galaxies brighter than limiting luminosity $L_{lim}$ is
\begin{equation}
N_{merg} = n(z)\ f_m \ T_m^{-1}
\end{equation}
where $n(z)$ is the co-moving number density of galaxies at redshift $z$ brighter than $L_{lim}$, 
$f_m$ is the fraction of merger candidates brighter than $L_{lim}$, 
and $T_m$ is the timescale during which merging galaxies can identified morphologically.  
It is important to note that the dynamical timescale for the entire merger process is
generally longer than the timescale during which merging systems can identified via
morphologies. $T_m$ may be different for different morphological measures ($G-M_{20}$,
asymmetry, visual classifications), and therefore may result in somewhat different
estimates of $N_{merg}$ if $T_m$ is not adjusted accordingly. 

For close pairs of galaxies within a given magnitude range, projected distance, and relative velocity, 
we adopt the formalism of Patton et al. (2000) where the galaxy merger rate is
\begin{equation}
N_{merg} = n(z) \ N_c(z)\ 0.5\ p(merg)\ T_p^{-1}
\end{equation}
where $n(z)$ is the co-moving number density of galaxies at redshift $z$ within the magnitude range of the observed
pairs, $N_c(z)$ is the average number of companions for galaxies within the observed magnitude range, the
0.5 factor accounts for the double counting of pairs, 
$p(merg)$ is the probability that the galaxy pair will merge (typically assumed to be 0.5), and $T_p$ is
the timescale for which merging galaxies appear as pairs
($\sim$ 0.25-0.5 Gyr for projected distances of $\leq$ 30 h$^{-1}$ kpc and relative velocities $\leq 500$ km s$^{-1}$; 
Lotz et al., in prep). 

To directly compare the morphological merger rate and the pair-count merger rate, one
must compare the pair-count merger rate over the same magnitude-range as the progenitor of the 
morphologically-disturbed sample.   We assume that mergers of mass ratios greater than 1:3 will be classified
as major mergers in $G$-$M_{20}$ and that the typical mass-to-light ratio is $\sim 1$.  
The kinematic galaxy pair studies by Patton et al. (2002) and Lin et al. (2004) draw both the primary and secondary galaxies
from the same magnitude range.  Therefore the progenitors of the $L_B > 0.4 L_B^*$ merger sample
are drawn from paired galaxies brighter than 0.1 $L_B^*$.  Adopting a Schechter function for the galaxy luminosity
function and neglecting the luminosity dependence of $N_c$, we find
\begin{eqnarray}
\begin{array}{ll}
N_{merg}& \sim \phi^*(z) \Gamma[2 + \alpha, 0.25 L_{lim}/L^*]\ N_c(z)\ 0.5\ p(merg)\ T_p^{-1} \\
        & = \phi^*(z) \Gamma[2+ \alpha,  L_{lim}/L*]\ f_m\ T_m^{-1}
\end{array}
\end{eqnarray}
and 
\begin{equation}
\frac{f_m}{N_c} \sim \frac{\Gamma[2+ \alpha, 0.25 L_{lim}/L*]}{\Gamma[2+ \alpha, L_{lim}/L*]}\ 0.5\ p(merg)\ \frac{T_m}{T_p}
\end{equation}
Assuming $\alpha = -1.3$, $p(merg) = 0.5$, and $L_{lim}= 0.4 L^*$, 
\begin{equation}
\frac{f_m}{N_c} \sim 0.29 \frac{T_m}{T_p}
\end{equation}
In the left-hand side of Figure 16, we plot $f_m$ for the EGS sample and 
$N_c$ for physical pairs with $\delta r \leq 30$ kpc, $\delta v \leq 500$ km s$^{-1}$,
and $-21 < M_B + Qz < -19$ where the luminosity evolution parameter Q=1 (Lin et al. 2004; Patton et al. 2002).
The $N_c$ values at $z \sim 0.7$ derived by Bell et al. (2006b) by applying the projected 
two-point correlation function 
to the COMBO-17 survey and at $0.5 < z < 1.5$ by Bundy et al. (2004) using near infrared-selected 
galaxy pairs are very
similar to those found by Lin et al. (2004).  The recent Millennium Galaxy Catalogue survey 
agrees with the Patton et al. (2002) 
$N_c$ value at z=0 (de Propris et al. 2007).  However, these studies have different selection criteria so we do not plot them in Figure 16. 

We find $f_m \sim N_c$, implying either $T_m \sim 3.4 T_p$
or that the fraction of morphologically-disturbed galaxies is larger than the merger rate implied by
the pair counts of bright galaxies. The latter could result from increasing $N_c$ at fainter magnitudes
(as seen in Fig. 2 of Lin et al. 2004).   Our morphological merger fraction could also be enhanced if many of
our merger candidates are minor mergers. Contamination of the morphological sample by irregular but not merging galaxies
would also artificially increase $f_m$. 
Both $T_m$ and $T_p$ are highly uncertain, although hydrodynamical simulations suggest
$T_m \sim 1.8-2.5 T_p$ (Lotz et al., in prep).

In the right-hand side of Figure 16, we compare our merger fraction to the literature values of the morphological merger fraction
(Le F\`{e}vre 2000, Conselice et al. 2003, Cassata et al. 2005, Scarlata et al. 2007a, de Propris et al. 2007). 
Our merger fractions are generally within the error-bars of these previous morphological studies at $z \leq 0.7$, but give 
somewhat lower values at $z \sim 1$. The exception is Kampczyk et al. (2007), who measured a very low fraction of 
visually-disturbed galaxies at $z \sim 0.7$ (2.4\%) but
estimate that they would detect only 1 in 5 mergers at $z \sim 0.7$ because of the low surface-brightnesses of tidal tails used
to classify mergers.  Because of this large correction factor and unknown uncertainties in their measured fractions, 
we have not plotted these points.  We do include the fraction of `irregulars' in the COSMOS field from the PCA analysis
of Scarlata et al. (2007a), although Kampczyk et al. (2007) suggest that only 30\% of these would be visually
classified as mergers. The Scarlata et al. (2007a) points are derived by integrating their best-fit Schechter functions for
the irregular and total galaxy populations down to our magnitude cutoff at $z=0.7$ ($M_B = 19.85$). 

The fraction of bright EGS galaxies classified as merger candidates in $G-M_{20}$ is $\sim$ 10\% .
This morphological major merger fraction does not evolve strongly for $0.2 < z < 1.2$ and is 
consistent with no evolution. Fitting a merger fraction evolution function $f_m(z) \propto (1+z)^m$ to the EGS
points, we find $m = 0.23 \pm 1.03$.   Including the literature values of the morphologically-determined merger fraction 
for $0.2 < z < 1.3$, we find $m = 1.26 \pm 0.56$.   If we include the $z \sim 0$ merger fraction
from de Propris et al. (2007) determined using asymmetries, the evolution in the merger fraction steepens to $m = 2.09 \pm 0.55$. 
We caution that including the merger fractions from the literature may not be valid given the variety of
morphological diagnostics and associated timescales needed to determine the merger fraction.  

The DEEP2 and COMBO-17 surveys find that the co-moving number density of
bright galaxies has not changed significantly since $z \sim 1.1$.  Integrating the Schechter function down
to 0.4 $L_*$ and assuming log10($\phi^*$)$=-2.5$ and $\alpha=-1.3$ (Faber et al. 2007), we find
the co-moving number density $n(z) = 2.1 \times 10^{-3}$ Mpc$^{-3}$.  Adopting $f_m = 0.10$ and $T_m \sim 0.5-1$ Gyr,   
we estimate that the merger rate for $L_B \ge 0.4 L^*_B$ is $2 - 4 \times 10^{-4}$\ Gyr$^{-1}$\ Mpc$^{-3}$ at $0.2 < z < 1.2$.  
Integrating this rate from $z = 1.1$ implies that 45-90\% of $L_B > 0.4 L^*_B$ galaxies at $z = 0.3$  
had a major merger in the previous 4.7 Gyr.   If many of our merger candidates have mass ratios less than 1:3, the
implied mass accretion rate would be lower but the evolution in the merger rate would remain unchanged. 

\begin{figure*}
\plotone{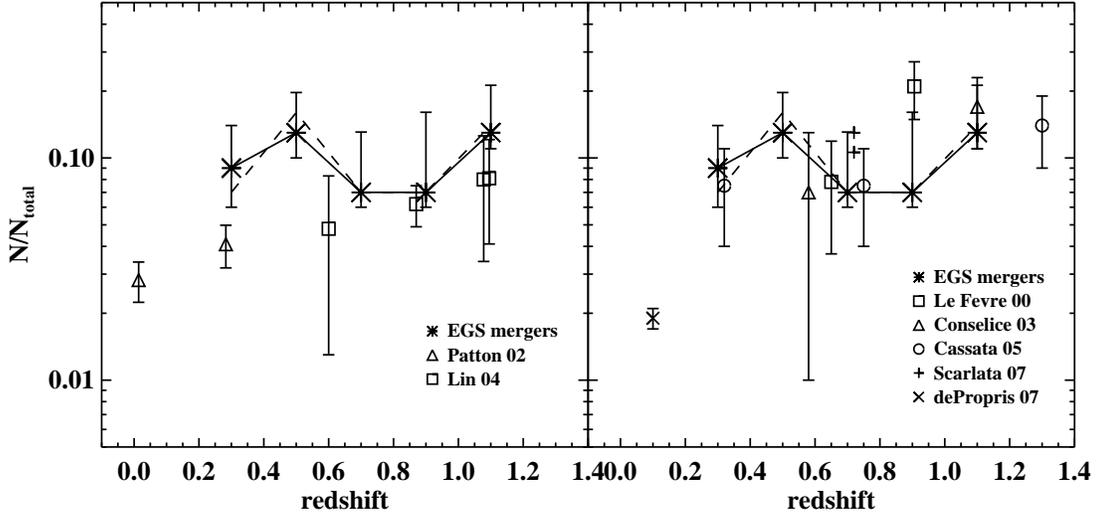}
\caption{The observed EGS morphological merger fractions v. literature values.  
$N_c$, the average number of companions from pair counts (Patton et al. 2002, Lin et al. 2004), are shown on the left.  
The morphological merger fraction from the 
literature for visual classifications (Le F\`{e}vre et al. 2000), rotational asymmetries (Conselice et al. 2003, Cassata et al. 
2005, de Propris et al. 2007),  and PCA analysis (Scarlata et al. 2007a) are shown on the right.  The Scarlata points are from integrated
the best fit Schechter functions to the irregular galaxies 
with (upper point) and without (lower point) corrections for the photometric redshifts.}
\end{figure*}

\section{Discussion}

The weak evolution of the merger fraction observed in the EGS at $0.2 < z < 1.2$  
is in disagreement  with previous claims for a dramatic increase in the $z \leq 1$ morphological merger rate with $m > 2$
(e.g., Brinchmann et al. 1998; Le F\`{e}vre et al. 2000; Conselice et al. 2003; Cassata et al. 2005; Kampczyk et al. 2007), 
despite the general agreement of the morphologically-determined merger fraction values at $z \sim 0.5 -1$ (Figure 16). 
Most of the evidence for strong evolution in the merger fraction comes from
the lowest and the highest redshift points (Figure 16).  At $z > 1.2$, the effects of bandpass shifting makes
measuring the merger fraction from morphologies difficult without high resolution near-infrared data. 
Both Brinchmann et al. (1998) and Le F\`{e}vre et al. (2000) visually classified galaxies from 
the Canada-France Redshift Survey and Autofib-Low Dispersion Spectrograph Survey using only
$HST$ $WFPC$ $I$-band images. A later study of mergers based on a similar visual classification scheme 
found much weaker evolution in
the merger rate when the morphological-dependence on wavelength was taken into account (Bundy et al. 2005).
The Cassata et al. (2005) claim for a strong increase in the merger fraction with redshift comes largely from 
galaxies at $z > 1.2$ where they can only measure the rest-frame ultraviolet morphologies. 
Only Conselice et al. (2003) has determined the merger fraction via morphologies using near-infrared
images at $z > 1$; this study used $HST$ NIC3 images of the Hubble Deep Field North and hence was limited
by small sample size.   Future observations with Wide Field Camera 3 on $HST$ should yield much larger 
high resolution near-infrared surveys for morphological studies at $z > 1.2$. 

Only a few works have attempted to measure the local merger fraction from morphologies (Kampczyk et al. 2007, 
de Propris et al. 2007) using visual classifications and asymmetries.  These are consistent with local
measurements of the kinematic pair fraction which also give lower merger rates than
the $z \sim 1$ measurements (Patton et al. 2002).  
Studies of the COSMOS ACS survey by Kampczyk et al. (2007) and Scarlata et al. (2007a) find evidence for
an increase in visually disturbed (Kampczyk et al. 2007) and quantitatively irregular (Scarlata et al. 2007a) galaxies at
$z \sim 0.7$.  However, because most of the COSMOS field is imaged in a single band-pass ($I$), galaxies
can be classified in rest-frame $B$ from the ACS data for only a limited redshift range.  
Both of these studies use a local sample of SDSS galaxies to determine the merger/irregular galaxy fraction at low redshift. 
Close pair studies using the DEEP1 survey (Bundy et al. 2004) and  the DEEP2 survey (Lin et al. 2004)
and the two-point correlation function from the COMBO-17 survey (Bell et al. 2006b) also find weak evolution with $m \sim 0.5$ 
in the pair fraction at $0.2 < z < 1.2$.  Kartaltepe et al. (2007) claim strong evolution in the fraction of close pairs in 
COSMOS identified via photometric redshifts only.  However, again most of their evidence for strong evolution comes
from the $z \sim 0$ SDSS pairs catalog of Allam et al. (2004); the evolution in the COSMOS $0.15 < z_{phot} < 1.05$ pairs
is much weaker.   The weak evolution in the pair counts and morphological disturbed galaxies at $0.2 < z < 1.2$ and
the low merger fraction observed at $z < 0.2$ suggest a rapid increase in the merger rate from $z \sim 0$ to $z \sim 0.2$. 
Clearly, more work is needed at $z < 0.2$ and at $z > 1$ to constrain the 
evolution in the galaxy merger rate.

The theoretical predictions for the galaxy merger rate over this epoch have changed in recent years.
Although N-body simulations of galaxy halos predict a halo merger fraction $\propto (1+z)^3$ (Gottl\"{o}ber,
Klypin, \& Kravtsov 2001), the {\it galaxy}
merger rate may evolve less dramatically than the halo merger rate because multiple galaxies 
occupy the same halo at late times (Berrier
et al. 2006). Interestingly, many semi-analytical models do not predict a dramatic increase 
in the fraction of merging galaxies brighter than 
$M_B = -20$ from $z \sim 0$ to $z \sim 1$ (Bell et al. 2006a; Benson et al. 2002).   The hydrodynamical
simulations analyzed by
Maller et al. (2006)  imply that the galaxy merger rate roughly doubles from $z=0$ to $z=0.6$.  However, they
find that $\sim$ 45\% of massive galaxies have experienced mergers with mass ratios greater than 1:4 
since $z=1$, in rough agreement with the merger rate we find here.  

\begin{figure*}
\plotone{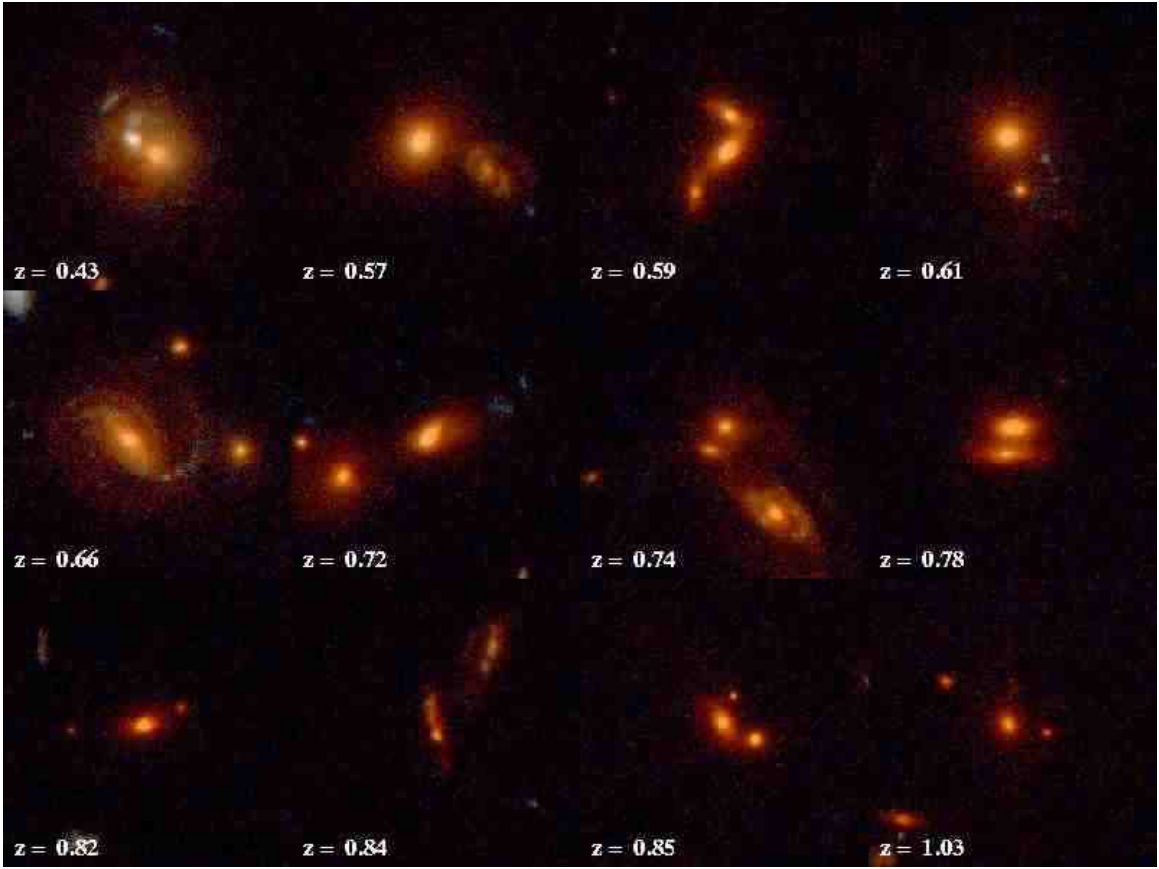}
\caption{ A subset of the red merger candidates; many are spheroidal-spheroidal or disk-spheroidal mergers.  
Each cutout is 9\arcsec\ wide.  The redshifts are given in the lower left corners.}
\end{figure*}

\begin{figure}
\plotone{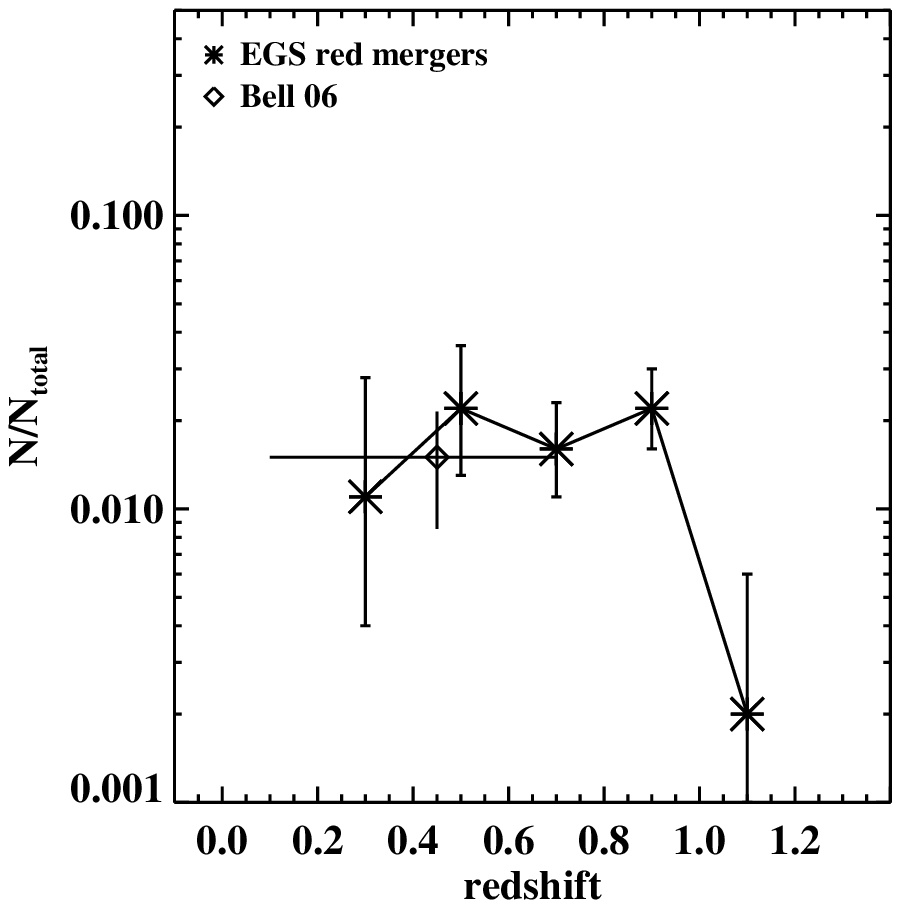}
\caption{ The fraction of EGS galaxies which are red mergers. For comparison,  
the disappationless merger rate from Bell et al. (2006a) is plotted.}
\end{figure}

We contradict previous $HST$ studies, which concluded that the increase in the faint
blue galaxy population and the volume-averaged star-formation density  ${\dot{\rho}}_*$ 
from $z \sim 0$ to 1 was driven by  major merger triggered star-formation in an increasing
population of peculiar galaxies 
(e.g., Glazebrook et al. 1995;  Le F\`{e}vre et al. 2000; Bridge et al. 2007).
Using the $G-M_{20}$ classification scheme, we are able to distinguish between late-type 
spirals/irregulars and major merger candidates (LPM04).  While the major merger fraction remains
constant from $0.4 < z < 1.2$, we find that the fraction of Sb-Ir
increases with redshift. These galaxies also dominate the infrared luminous galaxy
sample, which is the major contributor to star-production at $z \sim 1$.  
As concluded by other authors (Flores et al. 1999; 
Bell et al. 2005; Wolf et al. 2005; Melbourne et al. 2005; Menanteau et al. 2006, Noeske 
et al. 2007), the increase in star-formation density at $z\sim 1$ is not driven by merger-induced star-formation, but
rather by increased star-formation in the late-type disk population.  The corresponding increase in the fraction of bright 
late-types with redshift suggests that the increase in ${\dot{\rho}}_*$ is tied to disk gas-consumption and higher
gas fractions at $z \sim 1$ (e.g. Noeske et al. 2007).  However, we note that galaxies with the highest specific 
infrared luminosities (L(IR)$/$stellar mass) at $z \sim 1$ are more
likely to be in close kinematic pairs or morphologically-disturbed (Lin et al. 2007), and that LIRGs appear to
have higher asymmetries than typical spiral galaxies (Shi et al. 2006; Bridge et al. 2007).  Finally,  it is possible
that merging galaxies are mostly easily identified morphologically during the final merger while 
the peak in the star-formation rate/IR luminosity occurs during the first encounter before the interacting galaxies
become morphologically disturbed (Lotz et al., in prep). 

We find that the morphological makeup of the red sequence changes with redshift, 
with red Sb-Ir contributing roughly a third of the red sequence at $z \sim 1.1$,
while E/S0/Sa makeup $>$ 90\% of red galaxies at $z \sim 0.3$.  Our red galaxy morphology fractions 
agree with the visual
studies of $z \sim 0.7$ red galaxy morphology by Bell et al. (2005) and $z \leq 1$ galaxies by 
Weiner et al. (2005) (but see McIntosh et al. 2005).
We also find the declining contribution of spheroidals to the high-redshift red galaxy population suggested by 
visual morphological studies of extremely red objects at $z \sim 1.2$ (Moustakas et al. 2004) 
and $z \sim 2$ (Daddi et al. 2004).  Scarlata et al. (2007b) find similar evolution in the fraction of red galaxies with
quantitatively early-type morphologies, but have systematically lower fractions of early-types on the red sequence
(45\% at $z = 0.7$ and 60\% at $z = 0.3$).  Given that the number density of red
galaxies has increased by a factor of 2-4 from $z \sim 1.1$ to $z \sim 0.3$ (Faber et al. 2007, Brown et al.
2006, Bell et al. 2004a), the evolution of the morphological composition of the 
red sequence is consistent with a roughly constant number
density of red Sb-Ir and an increasing number density of red E/S0/Sa. 
Much of the evolution in the E/S0/Sa population may occur in fainter and less massive
systems (e.g. Bundy et al. 2005; Treu et al. 2005).
We find that bright red spheroidals are in place at $z \sim 1.1$, while red disks populate the 
faint end of the red sequence at $z \sim 1.1$ (Figure 13).  By $z \sim 0.3$, almost all of the faint red galaxies 
are spheroidals.  

Are these new red E/S0/Sa formed in mergers?  Although the merger rate does not increase strongly with redshift at $z < 1.2$, 
the observed merger fraction implies that 45-90\% of galaxies brighter than $0.4 L_B^*$ at $z \sim 0.3$ 
have undergone a major merger since $z \sim 1.1$.  The fraction of galaxies with E/S0/Sa morphologies
has roughly doubled over this redshift range. Assuming that the total number density of bright galaxies 
has not changed significantly since $z \sim 1.2$ (Faber et al. 2007), this implies that the
number density of E/S0/Sa has also doubled. If every major merger transformed two disk-dominated galaxies
into a spheroidal, we would expect the increase in E/S0/Sa to follow the integrated merger rate.  
The observed increase in the fraction of E/S0/Sa is a factor of 2-4 less than the expected increase in the merger 
remnant population, although marginally consistent within the uncertainties.  

Some mergers may not follow the standard Toomre (1977) picture of two 
gas-rich disks transforming into a gas-poor elliptical.  Our results suggest that at least half of the time 
the merging process does not dramatically transform the galaxies' morphologies. This conclusion is supported by recent 
work which finds that the end product of some gas-rich mergers can be disk-dominated 
(Springel \& Hernquist 2005; Hammer et al. 2005; Robertson et al. 2006) 
while spheroidal-spheroidal mergers will produce spheroidals (Bell et al. 2006a).   Fourteen
percent of the observed mergers lie on the red sequence.  
While some of these may be dusty or post-starburst merger remnants, some are clearly spheroidal-spheroidal
or disk-spheroidal mergers (Figure 17). In Figure 18, we plot the fraction of galaxies classified as
mergers on the red sequence. Our results give a low red merger fraction ($<$ 2\%)
and agree well with the dissipation-less merger fraction of Bell et al. (2006a).
Another possibility is that many of our merger candidates are minor merger events which do not
result in dramatic morphological transformations. 

It is tempting to directly associate the declining fraction of bright late-type spirals with the 
build-up of new E/S0/Sa via mergers.  While the faded remnants of the merger of bright blue disks 
cannot produce the brightest red spheroidals, they may produce faint red spheroidals. The recent production of
low mass spheroidals is consistent with evolution of the spheroidal stellar-mass function (Bundy et al. 2005)
and the indication that $M<10^{11}M_{\odot}$ ellipticals may have formed up to 40\% of their stars at $z < 1.2$ (Treu et al. 2005). 
Bright blue galaxies are in over-dense environments at $z \sim 1$ (Cooper et al. 2006), and hence may be more likely
to merge and form present-day spheroidals.  However, given that 
many of the brightest blue galaxies appear to be relatively undisturbed LIRGs, the transition from bright blue disk
to faint red spheroidal could happen via less violent processes such as gas consumption,  
AGN quenching, gas-stripping or minor mergers (e.g. Noeske et al. 2007,  Bell et al. 2005, Bundy et al. 2007). 
A better understanding of the triggering mechanisms for LIRGs, the evolution in the number density of faint spheroidals
and mergers, and the role of mergers in transforming galaxy morphology are needed to resolve this question. 

\section{Summary}
We have measured the quantitative morphologies $G$ and $M_{20}$ of $\sim 10,000$ relatively high
surface brightness galaxies
from $HST$ ACS imaging in $V_{F606W}$ and $I_{F814W}$ in $\sim$ 710 arcmin$^{2}$ of the Extended Groth Strip. 
Galaxies are classified as E/S0/Sa, Sb-Ir, and major mergers based on their rest-frame $B$-band
morphologies and local galaxy calibration of $G$ and $M_{20}$.   We examine the evolution of the morphological
fractions with redshift for a volume-limited sample of 3009 galaxies with $L_B > 0.4 L_B^*$ 
that assumes that galaxies fade by 1.3 $M_B$ per unit redshift.
We also examine the morphologies of 515 $0.2 < z < 1.2$ {\it Spitzer} MIPS 24 $\mu$m-selected 
L(IR) $> 10^{11} L_{\odot}$ galaxies in the EGS. 

(1) We find that the fraction of merger candidates is $\sim$ 10$\pm$2\% and 
does not evolve strongly from $z \sim 1.2$ to $z \sim 0.2$.   The fraction of E/S0/Sa has increased
by a factor of $\sim$ 2 from 21$\pm$3\% at $z \sim 1.1$ to 44$\pm$9\% at $z \sim 0.3$.  
The fraction of Sb-Ir has declined from 64$\pm$6\% to 47$\pm$9\% over the same redshift range. 

(2) The majority of LIRGs, which dominate the volume-averaged star-formation density at $z \sim 1$, are 
classified as Sb-Ir galaxies.  Only 15\% of these galaxies are classified as on-going mergers, 
suggesting that merger-driven star formation is not responsible for the bulk of star formation at $z \sim 1$. 

(3) The $G$-$M_{20}$ morphological classifications correlate strongly with rest-frame $U-B$ color.  Over 
75\% of E/S0/Sa are red, while 86\% of Sb-Ir  and 80\% of merger candidates are blue. 
Merger candidates are more likely to be blue than the
overall sample, with 14\% of mergers lying on the red sequence.

(4) The morphological makeup of the red sequence has changed since $z \sim 1.1$  where disk galaxies
are a third of red galaxies, while $\sim$ 90\% of $z \sim 0.3$ red galaxies are spheroidals.  Adopting the
number density evolution in red galaxies found by Faber et al. (2007), this implies that the build up of the
red sequence can be explained by the formation of faint red E/S0/Sa at late times.  

(5) The implied merger rate  at $0.2 < z < 1.2$ is $2 - 4 \times 10^{-4}$\ Gyr$^{-1}$\ Mpc$^{-3}$, 
with 45-90\% of $L > 0.4 L_B^*$ galaxies undergoing a merger between
$z \sim 1.1$ and $z \sim 0.3$.   While this rate is more than sufficient to produce the 
observed increase in the E/S0/Sa population, some mergers may not change galaxies morphological classifications.  
Fourteen percent of our merger candidates are red, and
some of these appear to be dissipationless mergers.  The disappearance of bright blue Sb-Ir galaxies
is consistent with the transformation of late-type disks into red spheroidals via major mergers.  However, 
many of the brightest blue galaxies are relatively undisturbed LIRGs.   More work on the triggering and/or
quenching mechanism for LIRGs and the transformation of galaxy morphology via mergers is needed to constrain the
formation mechanism for E/S0/Sa at $z < 1.2$.

We wish to thank D. Patton and T. Favorolo for their contributions to this work, and J. Berrier and
J. Bullock for access to their early manuscript.  We thank O. Ilbert for publicly releasing the
CFHTLS photometric redshift catalog.   We also thank E. Bell and D. McIntosh for useful discussions. JML would
like to thank P. Madau for support during the duration of this project, and acknowledges support from 
NASA grant NAG5-11513, NASA grants HST-G0-10134.13-A and HST-AR-10675-01-A from the Space Telescope Science Institute which is
operated by the AURA, Inc., under NASA contract NAS5-26555, Calspace grant to D.C.K., and the NOAO Leo Goldberg Fellowship.
AMH acknowledges support provided by  the Australian Research Council in the form of a QEII Fellowship (DP0557850). 
A.L.C. and J.A.N. are supported by NASA through Hubble Fellowship grants HF-01182.01-A and HST-HF-01165.01-A, awarded
by the Space Telescope Science Institute, which is operated by the Association of Universities for Research in Astronomy, Inc., 
for NASA, under contract NAS 5-26555. AJM appreciates support from the the National Science Foundation 
from grant AST-0302153 through the NSF Astronomy and Astrophysics Postdoctoral Fellows program.
ELF and CP acknowledge support by NASA through the {\it Spitzer Space Telescope} Fellowship Program through
 a contract issued by the Jet Propulsion Laboratory, California Institute of Technology under a contract with NASA.
Partial support was also provided through contract 1255094 from JPL/Caltech to the University of Arizona.

Based on observations made with the NASA/ESA {\it Hubble Space Telescope}, obtained from the data archive at the 
Space Telescope Science Institute. STScI is operated by the Association of Universities for Research in Astronomy, Inc. 
under NASA contract NAS 5-26555.  This work is based in part on observations made with the {it Spitzer Space Telescope}, which is 
operated by the Jet Propulsion Laboratory, California Institute of Technology under a contract with NASA. Support for this 
work was provided by NASA through an award issued by JPL/Caltech.

The DEEP2 survey is supported by NSF grants AST00-71198, AST00-71048, AST05-07428, and AST05-7483.
The DEIMOS spectrograph was funded by a grant from CARA (Keck Observatory), an NSF Facilities and Infrastructure grant (AST92-2540),
the Center for Particle Astrophysics and by gifts from Sun Microsystems and the Quantum Corporation.
The data presented herein were obtained at the W.M. Keck Observatory, which is
operated as a scientific partnership among the California Institute of
Technology, the University of California and the National Aeronautics
and Space Administration. The Observatory was made possible by the
generous financial support of the W.M. Keck Foundation. 

Based on observations obtained with MegaPrime/MegaCam, a joint project of CFHT and CEA/DAPNIA, 
at the Canada-France-Hawaii Telescope (CFHT) which is operated by the National Research Council (NRC) 
of Canada, the Institut National des Science de l'Univers of the Centre National de la Recherche Scientifique 
(CNRS) of France, and the University of Hawaii. This work is based in part on data products produced at the
 Canadian Astronomy Data Centre as part of the Canada-France-Hawaii Telescope Legacy Survey, a collaborative 
project of NRC and CNRS. 

We also wish to recognize and acknowledge the highly significant cultural role and reverence that the 
summit of Mauna Kea has always had within the indigenous Hawaiian community. We are most fortunate to have the 
opportunity to conduct observations from this mountain.

\clearpage
\begin{landscape}
  \LongTables
   \begin{deluxetable*}{lcclcclcclcclc}
   \tablewidth{0pc} 
   \tablecolumns{14} 
   \tablecaption{Extended Groth Strip $0.2 < z< 1.2$ Morphology Fractions} 
   \tablehead{\colhead{z} & \colhead{N(tot)} &  \colhead{N(E-Sa)} &  \colhead{f(E-Sa)\tablenotemark{a}} &  \colhead{f$_s$(E-Sa)\tablenotemark{b}} &  \colhead{N(Sb-Ir)} &  \colhead{f(Sb-Ir)\tablenotemark{a}} &  \colhead{f$_s$(Sb-Ir)\tablenotemark{b}} &  \colhead{N(M)} &  \colhead{f(M)\tablenotemark{a}} &  \colhead{f$_s$(M)\tablenotemark{b}}  &  \colhead{N(C)} &  \colhead{f(C)\tablenotemark{c}} &  \colhead{f$_s$(C)\tablenotemark{b}}} 
\startdata  
\cutinhead{$L_B > 0.4 L^*_B$ with completeness cut }  
 0.3  &   181 &    79 &   0.44 $^{+ 0.09}_{- 0.08}$ $-$ 0.04 & 0.44  &    86  & 0.47  $^{+ 0.09}_{- 0.08}$ $-$ 0.01 & 0.49  & 15 & 0.09 $^{+ 0.04}_{- 0.03}$ $+$ 0.03 & 0.07  &  1  &   0.01 $^{+ 0.014 }_{- 0.005 }$ &   0.00  \\ 
 0.5  &   312 &   104 &   0.33 $\pm 0.05$ $+$ 0.04            & 0.34  &   161  & 0.52  $\pm 0.07$ $+$ 0.02           & 0.49  & 45 & 0.13 $\pm 0.03 $ $+$ 0.06            & 0.16  &  2  &   0.01 $^{+ 0.009 }_{- 0.004 }$ &   0.00  \\ 
 0.7  &   765 &   291 &   0.37 $^{+ 0.04}_{- 0.03}$ $-$ 0.05 & 0.36  &   413  & 0.55  $^{+ 0.05}_{- 0.04}$ $-$ 0.06 & 0.55  & 56 & 0.07 $\pm 0.01 $ $+$ 0.06            & 0.07  &  5  &   0.01 $^{+ 0.005 }_{- 0.003 }$ &   0.00  \\ 
 0.9  &   676 &   227 &   0.32 $^{+ 0.04}_{- 0.03}$ $-$ 0.03 & 0.29  &   399  & 0.61  $\pm 0.05 $ $-$ 0.05          & 0.62  & 46 & 0.07 $\pm 0.01 $ $+$ 0.09            & 0.07  &  4  &   0.01 $^{+ 0.005 }_{- 0.003 }$ &   0.00  \\ 
 1.1  &   603 &   136 &   0.21 $\pm 0.03$ $-$ 0.00           & 0.15  &   374  & 0.64  $\pm 0.06$ $-$ 0.09           & 0.69  & 79 & 0.13 $\pm 0.02 $ $+$ 0.08            & 0.14  & 14  &   0.02 $^{+ 0.009 }_{- 0.007 }$ &   0.01  \\ 
\cutinhead{$L_B > 0.4 L^*_B$ with no completeness cut }  
 0.3  &   239 &    84 &   0.35 $^{+ 0.07}_{- 0.06}$ $-$0.02 & 0.35  & 132  & 0.55  $^{+ 0.09}_{- 0.08}$ $+$ 0.01 & 0.58  & 22 & 0.09 $^{+ 0.03}_{- 0.02}$ $+$ 0.04 & 0.08  &  1  & 0.00 $^{+ 0.011 }_{- 0.004 }$ &   0.00  \\ 
 0.5  &   400 &   111 &   0.27 $\pm 0.04$ $+$ 0.04          & 0.30  & 236  & 0.59  $^{+ 0.07}_{- 0.06}$ $+$ 0.02 & 0.55  & 51 & 0.12 $^{+ 0.03}_{- 0.02}$ $+$ 0.10 & 0.15  &  2  & 0.00 $^{+ 0.007 }_{- 0.003 }$ &   0.00  \\ 
 0.7  &   939 &   303 &   0.30 $\pm 0.03$ $-$ 0.03          & 0.32  & 559  & 0.60  $^{+ 0.05}_{- 0.04}$ $-$ 0.07 & 0.60  & 72 & 0.08 $\pm 0.01$ $+$ 0.07           & 0.07  &  5  & 0.01 $^{+ 0.004 }_{- 0.002 }$ &   0.00  \\ 
 0.9  &   782 &   233 &   0.27 $\pm 0.03$ $-$ 0.02          & 0.26  & 488  & 0.64  $\pm 0.05$ $-$ 0.04           & 0.65  & 57 & 0.07 $\pm 0.01$ $+$ 0.09           & 0.07  &  4  & 0.01 $^{+ 0.004 }_{- 0.003 }$ &   0.00  \\ 
 1.1  &   649 &   138 &   0.18 $\pm 0.03$ $+$ 0.00          & 0.14  & 407  & 0.64  $^{+ 0.06}_{- 0.05}$ $-$ 0.07 & 0.69  & 90 & 0.15 $\pm 0.02$ $+$ 0.08           & 0.15  & 14  & 0.02 $^{+ 0.009 }_{- 0.006 }$ &   0.01  \\ 
\cutinhead{$L_{IR} > 10^{11} L_{\odot}$}  
 0.3  &    14 &     1 &   0.06 $^{+ 0.25}_{- 0.06}$ $+$ 0.08&    0.17  &     8  &   0.54  $^{+ 0.59}_{- 0.29}$ $-$ 0.01 &    0.67  &     5 &   0.41 $^{+ 0.46}_{- 0.21}$ $-$ 0.04 &    0.17  &    0  &   \nodata &   \nodata  \\ 
 0.5  &    42 &    12 &   0.32 $^{+ 0.18}_{- 0.11}$ $+$ 0.01&    0.28  &    19  &   0.47  $^{+ 0.24}_{- 0.16}$ $+$ 0.11 &    0.38  &    11 &   0.23 $^{+ 0.17}_{- 0.11}$ $+$ 0.06 &    0.34  &    0  &   \nodata &   \nodata  \\ 
 0.7  &   151 &    34 &   0.22 $^{+ 0.07}_{- 0.05}$ $-$ 0.02&    0.22  &    96  &   0.63  $^{+ 0.13}_{- 0.11}$ $-$ 0.05 &    0.66  &    21 &   0.14 $^{+ 0.05}_{- 0.04}$ $+$ 0.03 &    0.11  &    0  &   \nodata &   \nodata  \\ 
 0.9  &   153 &    22 &   0.13 $^{+ 0.05}_{- 0.04}$ $+$ 0.00&    0.13  &   119  &   0.79  $^{+ 0.15}_{- 0.12}$ $-$ 0.07 &    0.81  &    12 &   0.08 $^{+ 0.04}_{- 0.03}$ $+$ 0.10 &    0.05  &    0  &   \nodata &   \nodata  \\ 
 1.1  &   144 &    12 &   0.07 $^{+ 0.04}_{- 0.03}$ $+$ 0.02&    0.05  &   111  &   0.77  $^{+ 0.15}_{- 0.13}$ $-$ 0.13 &    0.83  &    18 &   0.13 $^{+ 0.05}_{- 0.04}$ $+$ 0.08 &    0.10  &    3  &   0.02$^{+0.02}_{-0.01}$ &   0.01  \\ 
\enddata 
\tablecomments{ The morphology fractions are given for E/S0/Sa (E-Sa), Sb/Sc/Ir (Sb-Ir), merger candidates (M), and objects too compact to classify (C).}
\tablenotetext{a}{The fractions are corrected by the values given in Table 1. 
The first set of errors are the 67\% confidence intervals from Poisson statistics. 
Where $N<50$, the small number Poisson statistics are from Gehrels (1986). 
The second set of errors are the difference between the observed fraction and the fraction derived from 10,000 bootstrapped
realizations of morphological distributions.}
\tablenotetext{b}{The fraction derived for the spectroscopic redshift sub-sample and not corrected for
incompleteness.}
\tablenotetext{c}{ The errors are the 67\% confidence intervals from Poisson statistics. 
Where $N<50$, the small number Poisson statistics are from Gehrels (1986). }
\end{deluxetable*} 
\clearpage
\end{landscape}

\begin{deluxetable}{lcclclclcl} 
\tablewidth{0pc} 
\tablecolumns{10} 
\tabletypesize{\scriptsize} 
\tablecaption{Color-Dependent Morphology Fractions} 
\tablehead{\colhead{z} & \colhead{N(tot)} &  \colhead{N(E-Sa)} &  \colhead{f(E-Sa)\tablenotemark{a}}  &  \colhead{N(Sb-Ir)} &  \colhead{f(Sb-Ir)\tablenotemark{a}} &  \colhead{N(M)} &  \colhead{f(M)\tablenotemark{a}}  &  \colhead{N(C)} &  \colhead{f(C)\tablenotemark{a}} } 
\startdata  
\cutinhead{Red sequence galaxies }  
 0.3  &    66 &    61 &   0.92 $^{+ 0.08}_{- 0.21}$ &     2  &   0.03  $^{+ 0.05}_{- 0.02}$&     2 &   0.03 $^{+ 0.05}_{- 0.02}$   &    1  &   0.02 $^{+ 0.04}_{- 0.01}$  \\ 
 0.5  &    91 &    74 &   0.81 $^{+ 0.20}_{- 0.16}$ &    10  &   0.11  $^{+ 0.07}_{- 0.04}$&     7 &   0.08 $^{+ 0.06}_{- 0.03}$   &    0  &   \nodata  \\ 
 0.7  &   257 &   217 &   0.84 $^{+ 0.12}_{- 0.10}$ &    27  &   0.11  $^{+ 0.03}_{- 0.03}$&    12 &   0.05 $^{+ 0.02}_{- 0.02}$   &    2  &   0.01 $^{+ 0.01}_{- 0.005}$  \\ 
 0.9  &   217 &   166 &   0.76 $^{+ 0.12}_{- 0.10}$ &    36  &   0.17  $^{+ 0.05}_{- 0.04}$&    15 &   0.07 $^{+ 0.03}_{- 0.02}$   &    0  &   \nodata  \\ 
 1.1  &   137 &    95 &   0.69 $^{+ 0.14}_{- 0.12}$ &    40  &   0.29  $^{+ 0.09}_{- 0.07}$&     1 &   0.01 $^{+ 0.02}_{- 0.01}$   &    3  &   0.02 $^{+ 0.03}_{- 0.01}$  \\ 
\cutinhead{Green valley galaxies}  
 0.3  &     8 &     3 &   0.38 $^{+ 0.72}_{- 0.26}$ &     3  &   0.38  $^{+ 0.72}_{- 0.26}$&     1 &   0.12 $^{+ 0.51}_{- 0.11}$   &    0  &   \nodata  \\ 
 0.5  &    24 &    11 &   0.46 $^{+ 0.35}_{- 0.20}$ &     9  &   0.38  $^{+ 0.31}_{- 0.17}$&     4 &   0.17 $^{+ 0.21}_{- 0.10}$   &    0  &   \nodata  \\ 
 0.7  &    44 &    27 &   0.61 $^{+ 0.28}_{- 0.19}$ &    15  &   0.34  $^{+ 0.19}_{- 0.12}$&     2 &   0.05 $^{+ 0.08}_{- 0.03}$   &    1  &   0.02 $^{+ 0.07}_{- 0.02}$  \\ 
 0.9  &    53 &    19 &   0.36 $^{+ 0.18}_{- 0.11}$ &    29  &   0.55  $^{+ 0.23}_{- 0.15}$&     5 &   0.09 $^{+ 0.09}_{- 0.05}$   &    1  &   0.02 $^{+ 0.05}_{- 0.02}$  \\ 
 1.1  &    38 &    12 &   0.32 $^{+ 0.20}_{- 0.13}$ &    25  &   0.66  $^{+ 0.32}_{- 0.21}$&     0 &   0.00 $^{+ 0.06}_{- 0.00}$   &    1  &   0.03 $^{+ 0.08}_{- 0.02}$  \\ 
\cutinhead{Blue cloud galaxies}  
 0.3  &   108 &    15 &   0.14 $^{+ 0.07}_{- 0.04}$ &    81  &   0.75  $^{+ 0.17}_{- 0.14}$&    12 &   0.11 $^{+ 0.06}_{- 0.04}$   &    0  &   \nodata  \\ 
 0.5  &   197 &    19 &   0.10 $^{+ 0.04}_{- 0.03}$ &   142  &   0.72  $^{+ 0.12}_{- 0.10}$&    35 &   0.18 $^{+ 0.05}_{- 0.04}$   &    2  &   0.01 $^{+ 0.02}_{- 0.007 }$  \\ 
 0.7  &   467 &    47 &   0.10 $^{+ 0.02}_{- 0.02}$ &   371  &   0.79  $^{+ 0.08}_{- 0.07}$&    46 &   0.10 $^{+ 0.02}_{- 0.02}$   &    2  &   0.00 $^{+ 0.006}_{- 0.003 }$  \\ 
 0.9  &   415 &    42 &   0.10 $^{+ 0.02}_{- 0.02}$ &   334  &   0.80  $^{+ 0.09}_{- 0.08}$&    36 &   0.09 $^{+ 0.02}_{- 0.02}$   &    3  &   0.01 $^{+ 0.008}_{- 0.004 }$  \\ 
 1.1  &   423 &    28 &   0.07 $^{+ 0.02}_{- 0.01}$ &   309  &   0.73  $^{+ 0.08}_{- 0.07}$&    82 &   0.19 $^{+ 0.03}_{- 0.03}$   &   10  &   0.02 $^{+ 0.01}_{- 0.008 }$  \\ 
\enddata 
\tablecomments{ The morphology fractions are given for E/S0/Sa (E-Sa), Sb/Sc/Ir (Sb-Ir), merger candidates (M), and objects too compact to classify (C).}
\tablenotetext{a}{ The errors are the 67\% confidence intervals from Poisson statistics. 
Where $N<50$, the small number Poisson statistics are from Gehrels (1986). }
\end{deluxetable}

\end{document}